\documentclass[twocolumn, amssymb, amsmath, nobibnotes, aps, prx, superscriptaddress]{revtex4-1} 

\usepackage{graphicx}
\usepackage{float}
\usepackage{dcolumn}
\usepackage{bm}
\usepackage[colorlinks=true]{hyperref}
\hypersetup{
    colorlinks=true,       
    citecolor=blue,        
    urlcolor=blue           
} 
\usepackage{natbib}
\usepackage[caption=false]{subfig}

\begin{document}


\title{SYK Superconductivity: Quantum Kuramoto and Generalized Richardson Models}

\author{Hanteng Wang}
\affiliation{%
 School of Physics and Astronomy, University of Minnesota, Minneapolis, Minnesota 55455, USA
}%

\author{A. L. Chudnovskiy}
\affiliation{%
 1. Institut f\"ur Theoretische Physik, Universit\"at Hamburg,
Jungiusstr. 9, D-20355 Hamburg, Germany
}%

\author{Alexander Gorsky}

\affiliation{Institute for Information Transmission Problems,
Moscow, 127051, Russia}
\affiliation{Moscow Institute for Physics and Technology,
Dolgoprudnyi, 141700, Russia} 

\author{Alex Kamenev}

\affiliation{%
 School of Physics and Astronomy, University of Minnesota, Minneapolis, Minnesota 55455, USA
}%
\affiliation{%
 William I. Fine Theoretical Physics Institute, University of Minnesota, Minneapolis, Minnesota 55455, USA
}%

\date{\today}

\begin{abstract}
Sachdev-Ye-Kitaev (SYK) model has emerged as a new paradigm of the non-Fermi-liquid behavior. Here 
we investigate a possibility of having a superconducting off-diagonal long-range order (ODLRO) and a pseudogap phase within the SYK framework.
We found that ODLRO may be established in spin-1/2  version of the model with the time-reversal invariance and  
an extra attractive interaction. If the latter is taken as the on-site negative-$U$ Hubbard term, it leads to the pseudogap phase at $U<U_c$ dominated 
by quantum fluctuations of local phases.  These fluctuations are described  
by a quantum version of the Kuramoto model, traditionally employed to illustrate synchronization of classical non-linear oscillators. 
In the opposite limit of large $U$, the SYK+Hubbard model is approaching a certain generalization of the integrable Richardson 
model.  We present exact diagonalization studies, along with analytic solutions of the aforementioned limiting cases.
We also discuss possible holographic interpretations of the model, ODLRO and the pseudogap. 
\end{abstract}

\maketitle


\section{Introduction}
\label{sec:Intro}

Sachdev-Ye-Kitaev (SYK) model \cite{Sachdev1993,Kitaev2015} has received a great deal of attention in recent years as being an exactly solvable model with non-Fermi-liquid properties \cite{Song2017,davison2017thermoelectric,banerjee2017solvable,patel2018magnetotransport,chowdhury2018translationally,Feigelman2018,altland2019SYK}. It also admits a dual holographic description in terms of Jackiv-Teiteboum (JT) AdS$_2$ gravity \cite{Kitaev2015,CommentsSYK16,almheiri2015models,Sachdev2015PRX,engelsoy2016investigation,cotler2017black,Kitaev&Suh,Kourkoulou17}  and saturates the limiting rate \cite{maldacena2016bound,CommentsSYK16}  of chaotization  \cite{you2017sachdev,Jensen16,cotler2017chaos,garcia2018chaotic,gu2017local,gu2017energy,bi2017instability,sonner2017eigenstate,chen2017competition,gharibyan2018onset,altland2018quantum}. Although the initial SYK model is zero-dimensional (0D) with all-to-all random interactions, it was soon generalized to include $D$-dimensional arrays of connected SYK grains \cite{Song2017,gu2017local,Xu2017,Jian2017,zhang2017dispersive,chowdhury2018translationally,jian2018quantum,Shenoy2018,zhang2018topological,Xu2019,altland2019quantum}. Such models were shown to exhibit $T$-linear resistivity, making them attractive candidates for description of strongly correlated materials \cite{gurvitch1987resistivity}. 
An account of quantum fluctuations in such arrays reveals \cite{altland2019quantum} a quantum phase transition (QPT) between a gapless (thermal) insulator and the Fermi liquid at certain critical inter-grain coupling. In these picture the $T$-linear metallic phase appears as the quantum critical region \cite{SachdevBook} of the aforementioned QPT.

Success of the SYK model in describing the non Fermi liquid state raises the question if superconductivity may be included in the same framework. 
A number of models were suggested with this goal in mind both in  0D \cite{esterlis2019cooper,hauck2019eliashberg,wang2019solvable} and in the array \cite{patel2018coherent,chowdhury2019intrinsic,chowdhury2019unreasonable}  context.  All of them found that the original SYK model must be upgraded to complex spin-full fermions with an extra mechanism of attraction, such as phonons \cite{esterlis2019cooper,*hauck2019eliashberg,wang2019solvable}, pair hopping  \cite{patel2018coherent}, or special correlations between matrix elements \cite{chowdhury2019intrinsic,*chowdhury2019unreasonable}. Such upgraded  SYK-like models indeed exhibit superconducting correlations, which 
may be treated within the large $N$ mean-field approach. Similarly to the Fermi liquid BCS mechanism, an infinitesimal attraction is sufficient to develop the superconductivity.  

In this paper we consider a different 0D model, where the attraction is provided by a negative $U$ Hubbard term. Contrary to the mechanisms mentioned above, the mean-field treatment completely fails to describe the SYK+Hubbard model even in the $N\to \infty$ limit. This is because the Hubbard term does not inhibit on-site phase fluctuations, which invalidate the mean-field approach. Such quantum phase fluctuations  result in an insulating {\em pseudogap} phase at $U<U_c$, where $U_c$ is a critical attraction strength.  For $U>U_c$ there is a superconducting ``dome'' on the $U$ vs. temperature plane. The 
superconducting phase under the dome is also strongly affected by the quantum fluctuations and does not conform to the mean-filed description. 

In view of the failure of the mean-field, one needs to develop alternative means, capable of treating strongly fluctuating phases. Fortunately, within 
the 0D framework this can be  achieved. In the limit of large $U$ we found that the model may be mapped onto a certain generalization of the exactly solvable Richardson model  \cite{Richardson1963,vonDelft2000,Dukelsky2004}. It's solution reveals a superconducting low temperature state with the first order 
transition to the normal non Fermi liquid state at $T_c\propto U^{-1}$. The first order transition between a superconductor and a non Fermi liquid has been already noticed in Refs.~\cite{chubukov2003first,patel2018coherent}. It's possible that SYK+Hubbard and the associated Richardson models provide the 
simplest cartoon for this phenomenon.  

In the opposite limit of a weak attraction, the phase fluctuations may be described by an effective model, which we call the {\em quantum Kuramoto} model.
The classical Kuramoto model is a paradigm for synchronization of non-linear stochastic oscillators \cite{Kuramoto1975,daido1992quasientrainment,wiesenfeld1998frequency,strogatz2000kuramoto,acebron2005kuramoto,arenas2006synchronization,gomez2007synchronizability,dorfler2013synchronization,boccaletti2014structure,witthaut2017classical,DSouza2019explosive}. It's quantum counterpart  
provides a description of a continuous QPT between the pseudogap state with unsynchronized phases and the phase-coherent superconductor. We found it 
remarkable that the SYK framework is capable to exhibit  the pseudogap physics.   

To verify validity of this theory we resort to an exact diagonalization of spin-1/2 SYK+Hubbard model.  To detect superconductivity numerically in a finite size system, 
we employ the  notion of 
the off-diagonal long-range order (ODLRO) \cite{yang1962concept,Leggett2001}. It allows for a sharp definition of the condensate fraction and its dependence on temperature and the attraction strength for a large, but finite $N$ (number of sites).  Numerical results are in a qualitative (and in cases where numerical coefficients may be evaluated, a quantitative)  agreement with the theory.

The paper is organized in the following way. In section \ref{sec:Models} we discuss the models and the notations. Section \ref{sec:ODLRO} is devoted to the notion of ODLRO. It is followed by section \ref{sec:mean-field}, where we outline mean-field treatment and expectations for the models at hand, the latter are then compared with the results of the exact diagonalization in section \ref{sec:ED}. In section \ref{sec:quantum-fluctuations} we explain numerical observations by mapping onto Richardson and quantum Kuramoto models in the regimes of  strong and weak attraction correspondingly. In section  \ref{sec:holography} we discuss a possible  holographic  interpretation of the fluctuation-dominated SYK superconductivity in terms of the ``bulk'' description.
Finally, section  \ref{sec:conclusions} briefly summarizes our findings and lists some open problems.


\section{Notations and Models} 
\label{sec:Models}

We consider 0D models, consisting of $N\gg 1$ orbitals (or sites), labeled as $i,j,\ldots = 1,2,\ldots,N$. Each orbital may be occupied by a complex spin-$1/2$ fermion annihilated with the operator $c_{i\sigma}$, where $\sigma= \downarrow,\uparrow$ is the spin index. In the spirit of the SYK model, we assume that 
all orbitals are exactly degenerate with the on-site energy taken to be zero. The orbitals interact through the four-fermion interaction with {\em real} spin-independent matrix elements. These interactions are summarized by the SYK part of the Hamiltonian:    
\begin{equation} \label{eq:model}
  H_\mathrm{SYK} = {\frac{1}{2}}\sum_{ijkl; \sigma\sigma^{\prime}}^N {J_{ij;kl}  \big[  
  c_{i \sigma}^{\dagger} c_{j \sigma^{\prime}}^{\dagger} c_{k \sigma^{\prime}} c_{l \sigma} +
  c_{l \sigma}^{\dagger} c_{k \sigma^{\prime}}^{\dagger} c_{j \sigma^{\prime}} c_{i \sigma}
  \big]},  
\end{equation}
where $J_{ij;kl}$ is a real tensor with the following symmetry properties:
\begin{equation} \label{eq:J-symmetry}
J_{ij;kl} = -J_{ji;kl} =-J_{ij;lk}=J_{lk;ji}.   
\end{equation}
We also demand that non-zero elements must have all four indexes $i,j,k,l$  distinct. Up to these symmetries, the matrix elements 
$J_{ij;kl}$ are assumed to be real independent random variables, drawn from the Gaussian distribution with the zero mean, $\langle J_{ij;kl} \rangle=0$, 
and the variance    
\begin{equation}
\langle J_{ij;kl}^2 \rangle=J^2/(4N)^3.
\label{var_J}
\end{equation}   

We'll show below (both numerically and analytically) that the pure SYK Hamiltonian (\ref{eq:model}) does not lead to ODLRO~\cite{Tarnopolsky2020}. 
For ODLRO to develop, one needs to supplement SYK Hamiltonian with an attractive term, facilitating fermion pairing. One possibility is a site-local 
negative $U$ Hubbard term:  
\begin{equation} \label{eq:Hubbard}
  H_\mathrm{Hub} =  - U \sum_{i}^N{ c_{i \uparrow}^{\dagger} c_{i \downarrow}^{\dagger} c_{i \downarrow} c_{i \uparrow} } 
  -\mu \sum_{i,\sigma}^N{ c_{i \sigma}^{\dagger}  c_{i \sigma} }.
\end{equation}
Another option is all-to-all pair hopping \cite{patel2018coherent}:
\begin{equation} \label{eq:pair-hopping}
  H_\mathrm{p-hop} =  - {\frac{U}{N}} \sum_{ij}^N{ c_{i \uparrow}^{\dagger} c_{i \downarrow}^{\dagger} c_{j \downarrow} c_{j \uparrow} }
   -\mu \sum_{i,\sigma}^N{ c_{i \sigma}^{\dagger}  c_{i \sigma} },
\end{equation}
which annihilates a pair at an orbital $j$ and creates at, in general, different orbital $i$.    Both Hamiltonians contain a chemical potential to adjust 
the occupation fraction. The three Hamiltonians, written  above, conserve particle number and are symmetric under the time-reversal transformations.  States of these models are governed by temperature, $T$,  fermion occupation number, $N_f$, and the dimensionless parameter, $U/J$, characterizing the attraction strength.  

In the absence of the SYK term the ground state of the pure Hubbard model, Eq.~(\ref{eq:Hubbard}),  consists of localized pairs and does not exhibit ODLRO. Its energy is obviously $-U$ per fermion pair and its degeneracy is given by the number of combinatorial possibilities of distributing a given number of pairs among $N$ orbitals.
Excited states are formed by breaking some of the pairs and creating single occupied orbitals with zero energy. As we show below, ODLRO may be established, mediated by the SYK interactions.   

The pure pair-hopping Hamiltonian, Eq.~(\ref{eq:pair-hopping}), is somewhat different. It constitutes a limiting case of the Richardson model \cite{Richardson1963,vonDelft2000,Dukelsky2004} 
(see section \ref{sec:Richardson} and Appendix \ref{app:Richardson} for details). The latter predicts a non-degenerate ground state with ODLRO separated by the gap, $\propto U$, from the first excited state, which is $(N-1)$-fold degenerate. We'll show that SYK interactions do not destroy ODLRO, but weaken it substantially if $J>U$.

Numerically we first block diagonalize the  $2^{2N}\times 2^{2N}$ matrix Hamiltonian in the  many-body space, using particle number conservation and other 
symmetries (e.g. particle-hole symmetry for the half-filled case). We then exactly diagonalize the relevant blocks   
to extract their spectrum and eigenfunctions. 


\section{The off-diagonal long-range order}
\label{sec:ODLRO}

The standard definition of the superconductivity implies a finite anomalous expectation value,  
$\bar\Delta_i\propto \langle  c_{i \uparrow}^{\dagger} c_{i \downarrow}^{\dagger} \rangle$. It is clear however, that for a finite size system with a particle conserving Hamiltonian such expectation value is bound to vanish. 
One thus needs another measure of the superconducting order. The corresponding concept of ODLRO is well known from, e.g., the theory of cold atom Bose condensates in optical or magnetic traps \cite{Leggett2001}.

Let us define the bosonic pair creation operator as 
\begin{equation}  \label{eq:bosons}
 b_i^{\dagger}   =  c_{i \uparrow}^{\dagger} c_{i \downarrow}^{\dagger}. 
\end{equation}
Since there can't  be more than one such boson per orbital, we are dealing with the hard-core bosonic particles. One then defines the reduced {\em single-particle} bosonic density matrix as
\begin{equation}  \label{eq:density-matrix}
  \rho_{ij} = \langle b_i^{\dagger} b_j \rangle,
\end{equation}
where $\langle\ldots \rangle$ implies the exact many-body ground state (or thermal) expectation value. Defined this way, $\rho_{ij}$, is an $N\times N$ 
positive-definite matrix. Its trace is a total number of local pairs, which is less or equal than $N_f/2$ (we typically consider half-filled systems with $N_f=N$).
One is interested in the spectrum of eigenvalues of $\rho_{ij}$: $\lambda_{\alpha}$, where $\lambda_0  \geq \lambda_1 \geq ... \geq \lambda_{N-1}\geq 0$ and $\sum_{\alpha=0}^{N-1}\lambda_\alpha \leq N_f/2$. The absence of the pair condensate corresponds to all $N$ eigenvalues $\lambda_{\alpha}$ being of order one, $ O(1)$.    On the other hand, the pair condensate corresponds to the largest eigenvalue $\lambda_0$ being $O(N)$, while the remaining  $N-1$ eigenvalues being $O(1)$.     

\begin{figure}[htb]
  \centering
  \includegraphics[width=0.5\textwidth]{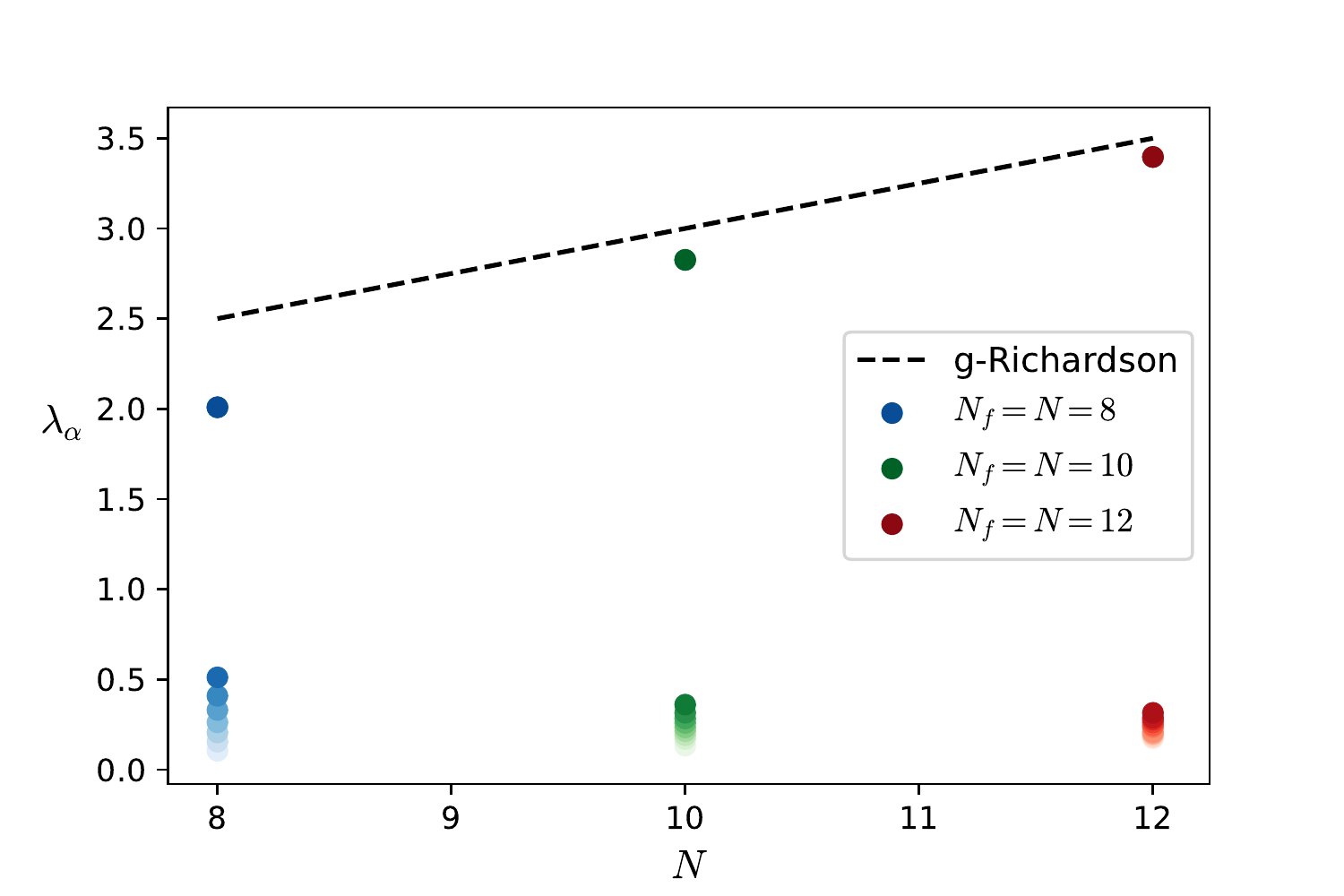}
  \caption{Spectrum of $\rho_{ij}$, i.e. $\lambda_{\alpha}$ vs. number of orbitals $N$ for the ground state of SYK+Hubbard model  with $U/J=2$ and $N_f=N$. Dashed line is a result from generalized Richardson model (Section \ref{sec:Richardson}). }
  \label{fig:ODLRO}
\end{figure}

\begin{figure}[htb]
  \centering
  \includegraphics[width=0.5\textwidth]{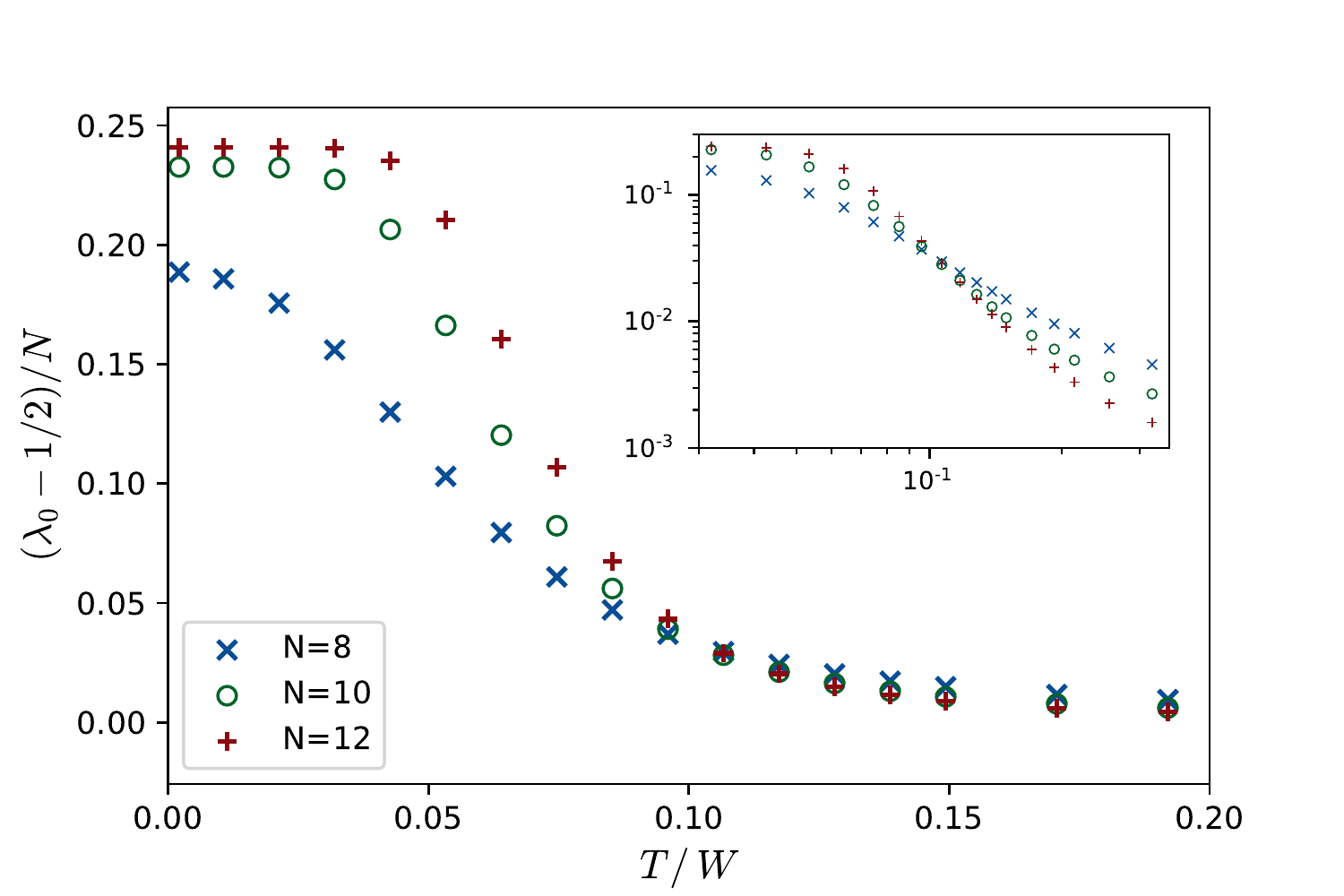}  
  \caption{$(\lambda_0-1/2)/N$ vs. temperature of SYK+Hubbard model with $U/J=2$. The temperature is 
  normalized to $W= 3J^2/32U$ (Section \ref{sec:Richardson}). Inset: vicinity of the crossing point.}
  \label{fig:T-ODLRO}
\end{figure}

Figure \ref{fig:ODLRO} shows  $T=0$ spectrum of $\rho_{ij}$ for SYK +Hubbard model  with $U/J=2$ for $N=N_f=8,10,12$. One can clearly see the largest eigenvalue splits from the rest and approaches $N/4+1/2$. The remaining eigenvalues coalesce towards $\approx 1/4$. This behavior may be understood with the help of the generalized Richardson model, as explained in  Section \ref{sec:Richardson}. The presence of the single eigenvalue with the $O(N)$ scaling is the hallmark of ODLRO \cite{Leggett2001}. Indeed, admitting   a nonzero anomalous average $\bar\Delta_i\propto \langle  c_{i \uparrow}^{\dagger} c_{i \downarrow}^{\dagger} \rangle$, one finds $\rho_{ij} \propto \bar\Delta_i \Delta_j $. This is the rank-1 matrix with the single non-zero eigenvalue, given by its trace ($\propto N$). 

Figure \ref{fig:T-ODLRO} shows temperature dependence of the condensate density, $(\lambda_0-1/2)/N$, (subtraction of $1/2$ is motivated by the expectation that, in the absence of ODLRO, all $\lambda$'s approach $1/2$). One notices the approximate crossing point at $T_c\approx  0.1W$, where $W$ is the energy scale of the Richardson model, $W= 3J^2/32U$, (see Eq.~(\ref{eq:Richardson}) in Section \ref{sec:Richardson}). 
Such crossing point indicates a phase transition in the $N\to \infty$ limit between phases with a finite and zero condensate density. 


\section{Mean-Field Treatment}
\label{sec:mean-field} 

To develop a large $N$ mean-field treatment, one follows the standard root \cite{Kitaev2015,Bagrets-Altland-Kamenev2016}  of averaging over the random SYK matrix elements 
$J_{ij;kl}$ and deriving the so-called $G\Sigma$ action. There is a peculiarity, though, associated with the matrix elements being real.
It is coming from the fact that there are two distinct terms in the square bracket on the right hand side of Eq.~(\ref{eq:model}), see Appendix \ref{app:mean-field}. 
Upon averaging over the Gaussian distribution of   $J_{ij;kl}$, one obtains two types of terms which are expressed through the normal and anomalous two-point fields: 
 \begin{eqnarray}  
 \nonumber && 
 G_{\tau,\tau'}\!=\! {-\frac{1}{N}} \sum_{i}^N c_{i\sigma}(\tau) c_{i\sigma}^\dagger(\tau'); \\ 
 && 
  F_{\tau,\tau'}\!=\! {-\frac{1}{N}} \sum_{i}^N  c_{i\downarrow}(\tau) c_{i\uparrow}(\tau'), \label{eq:GF}
\end{eqnarray} 
The normal component is spin-diagonal and independent of the spin-projection.  Here we have suppressed replica indexes for brevity. 
The normal and anomalous components may be combined in the Nambu matrix field 
$\hat G_{\tau,\tau'}$.  The definitions (\ref{eq:GF}) are enforced by conjugate non-local fields, which may be also combined into the Nambu space matrix  $\hat \Sigma_{\tau,\tau'}$,  playing the role of the self-energy. 

The Hubbard term, Eq.~(\ref{eq:Hubbard}), may be decoupled in the Cooper channel with the help of the local fields $\Delta_i(\tau)$, leading to the 
effective action of the form: 
\begin{eqnarray}
\hskip -.4cm S \! = && \sum_i^N \int\!\! d\tau   \left[ \frac{|\Delta_i|^2}{U} - \frac{1}{2}\, \mathrm{Tr}\ln(\partial_\tau +\mu - \hat\Sigma-\hat\Delta_i)\right]  \nonumber \\
&&\hskip -.6cm-\, N\int\!\!\! \int d\tau d\tau' \left[\hat\Sigma_{\tau,\tau'} \hat G_{\tau',\tau} + \frac{J^2}{64} \left( \bar{F}_{\tau\tau'}^2F_{\tau\tau'}^2  +
G_{\tau\tau'}^4\right) \right], 
							\label{eq:action}   
\end{eqnarray}
where $\hat\Delta_i=\Delta_i\sigma_+ +\bar\Delta_i\sigma_-$ is the off-diagonal Nambu matrix. For the pair-hopping model, Eq.~(\ref{eq:pair-hopping}), one needs a single field $\Delta(\tau)$ to decouple it. One thus arrives at the same action (\ref{eq:action}) with the constraint $\Delta_i =\Delta$. In the latter case 
there is a large factor $N$ in front of the entire action, justifying the mean-field saddle point approximation.    

The mean-field equations, obtained upon variation of the action over the matrix fields $\hat G$, $\hat\Sigma$ as well as over $\Delta$ are 
specified in Appendix \ref{app:mean-field}.  Their numerical analysis 
\cite{Tarnopolsky2020} shows that in the absence of attraction ($U=0$ and thus $\Delta_i=0$) the lowest free energy solution is purely 
normal, i.e. $F_{\tau,\tau'}=0$, while $\hat G_{\tau,\tau'}\propto |\tau-\tau'|^{-1/2}$, same as in conventional complex-$J$ SYK model.  

One can investigate now stability of such non-superconducting SYK solution against a small attractive $U$ perturbation. The corresponding self-consistency equation for $\Delta$ takes the form $U^{-1} = {\cal C}(\Delta)$, where the Cooper channel polarization ${\cal C}=\int d\tau G^2_\tau$. In the 
normal phase of SYK, $G_\tau\propto (J\tau)^{-1/2}$, and therefore ${\cal C}$ is given by the logarithmic integral. In the IR limit the latter is cut by either temperature or 
$|\Delta|$, leading to $U^{-1}\propto J^{-1}\ln(J/|\Delta|)$ and thus $|\Delta| \sim J e^{-\mathrm{const}\cdot J/U}$ for $U\ll J$. Thus the mean-field treatment predicts that, similarly to BCS case,  {\em an arbitrarily weak attraction} results in a finite superconducting order parameter, albeit an exponentially small one. 

A detailed calculation, presented in Appendix \ref{app:mean-field}, leads to the following mean-field solution for the absolute value of the order parameter 
\begin{equation}       \label{eq:mean-field} 
|\Delta|\propto \left\{\begin{array}{ll} 
J\,e^{- J\sqrt{\pi}/(8\sqrt{2} U)}; & \quad U\ll J,\\
U/2; &\quad  J\ll U.
\end{array}\right. 
\end{equation}
It is worth mentioning that the energy gap in the many-body spectrum scales as $|\Delta|^2/J$ for $U\ll J$ and as $|\Delta|$ for $U\gg J$, Appendix \ref{app:mean-field}. 
  
As mentioned above, one expects the mean-field treatment to be accurate for the SYK+pair hopping model in $N\to \infty$ limit.  It is not clear a priory if SYK+Hubbard is also accurately described by this theory. Indeed, in the latter case the order parameters, $\Delta_i$, on individual orbitals fluctuate independently (first line in Eq.~(\ref{eq:action})) and such fluctuations are not necessarily decreasing as $N\to \infty$.    
To check this we perform finite-size exact diagonalization study, summarized below.

\section{Exact Diagonalization} 
\label{sec:ED}

\begin{figure}[H]
  \centering
  \includegraphics[width=0.45\textwidth]{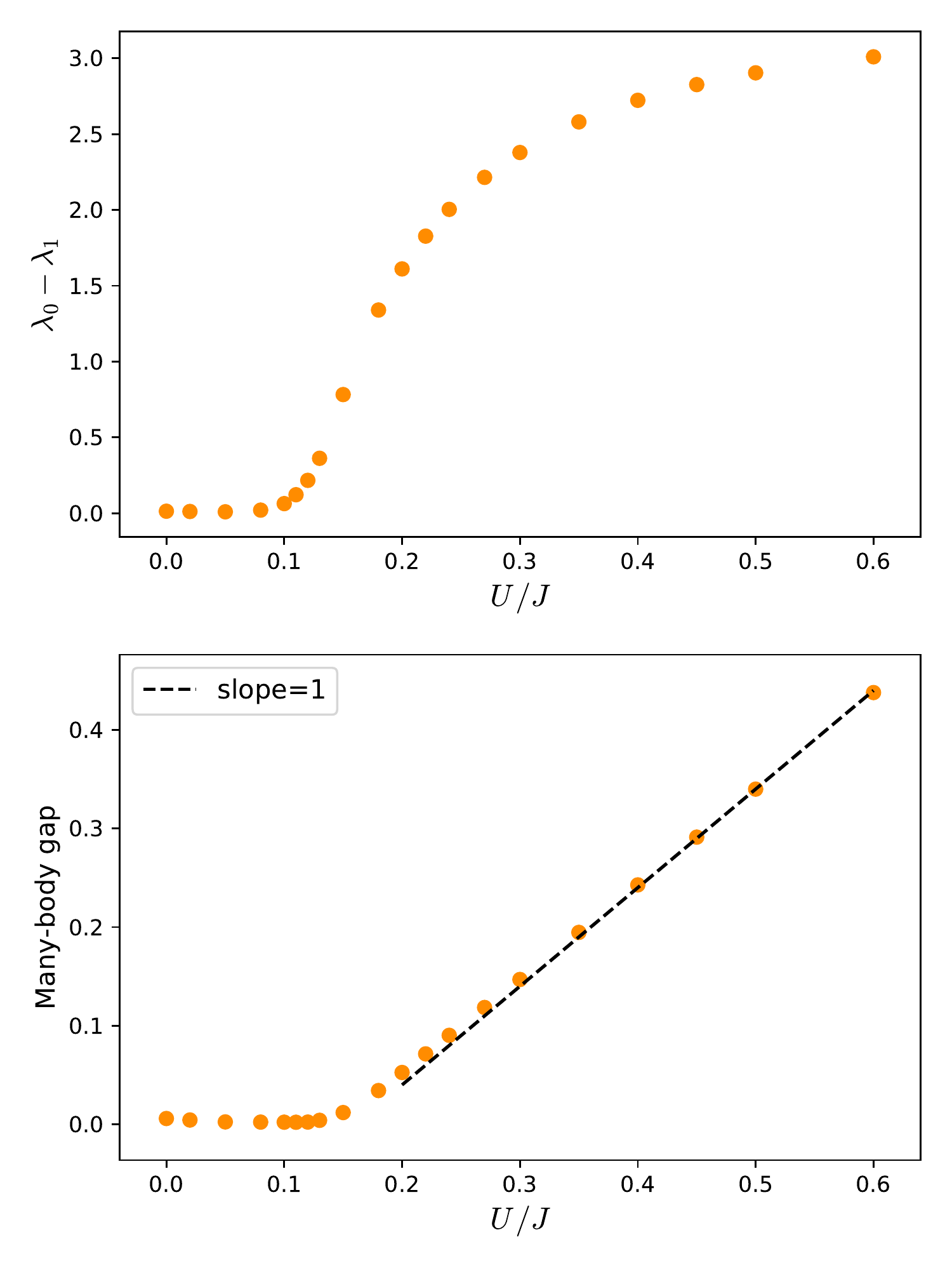} 
  \vspace{-0.5cm}
  \caption{ODLRO (top) and many-body energy gap in units of $J$ (bottom) vs. $U/J$ for the SYK+pair-hopping model with $N_f=N=12$.}
  \label{fig:pair-hopping-U}
\end{figure}

Figure \ref{fig:pair-hopping-U} shows the exact diagonalization results for the SYK+pair hopping Hamiltonian, Eqs.~(\ref{eq:model}), (\ref{eq:pair-hopping}), 
for the half-filled $N=12$ case -- the largest size accessible in our simulations. The top panel shows ODLRO, defined as the difference between the largest and the second largest eigenvalues of $\rho_{ij}$, Eq.~(\ref{eq:density-matrix}), as a function of $U/J$.  The bottom panel shows the gap in the many-body spectrum, defined as the difference between the energies of the first excited and the ground-state, also as a function of $U/J$. At $U\gg J$ the ODLRO saturates to $N/4$, while the many-body gap approaches $U$ - in agreement with the mean-field. Due to finite size effects, it is hard to draw definitive conclusions about small $U$ behavior. Qualitatively it is also consistent with  the mean-field expectations, Eq.~(\ref{eq:mean-field}).     

\begin{figure}[H]
  \centering
  \includegraphics[width=0.48\textwidth]{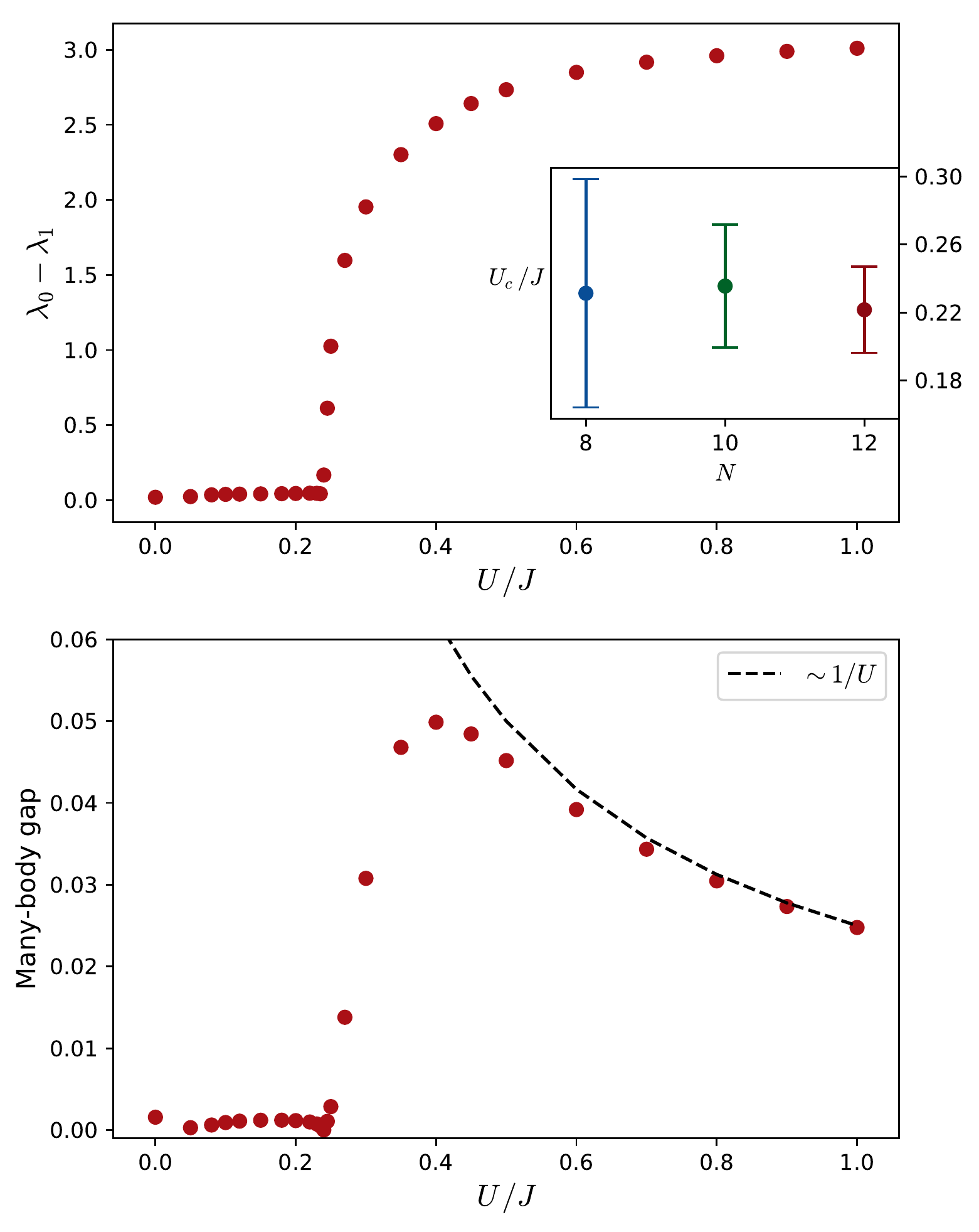}
  \vspace{-0.5cm}
  \caption{ODLRO (top) and many-body gap (bottom) vs. $U/J$ for the SYK+Hubbard model with $N_f=N=12$. Inset: $U_c$ vs. system size $N$. Error bars 
  reflect statistical fluctuations.}
  \label{fig:Hubbard-U}
\end{figure}

\begin{figure}[htb]
  \centering
  \includegraphics[width=0.5\textwidth]{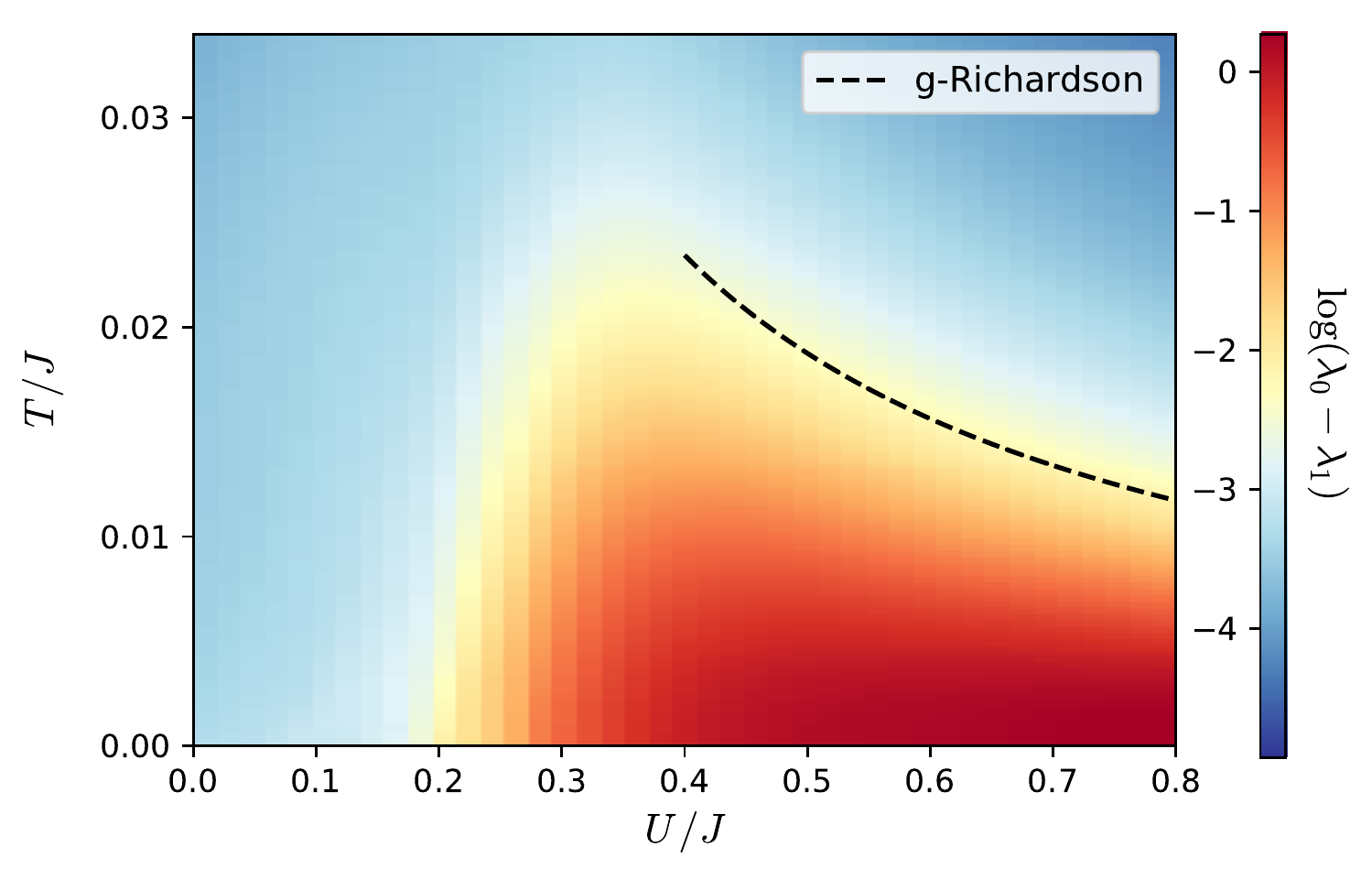}
  \caption{Superconducting ``dome''. Color plot of  $\log(\lambda_0-\lambda_1)$ for the SYK+Hubbard model with $N_f=N=8$  on $T$ vs. $U/J$ phase plane. The dashed line 
  is prediction for $T_c$ for the generalized Richardson model, section \ref{sec:Richardson}. }
  \label{fig:Phase_diagram}
\end{figure}

This behavior should be contrasted with the results of the exact diagonalization of the SYK+Hubbard, Eqs.~(\ref{eq:model}), (\ref{eq:Hubbard}), presented in  Fig.~\ref{fig:Hubbard-U}. One notices a critical value $U_c\approx 0.24 J$, below which there is no any evidence of neither ODLRO, nor the many-body gap (beyond a finite-size effect of the SYK model). As indicated in the inset, $U_c$ does not decrease with increasing $N$ and thus it's unlikely to be a finite-size artifact. Another marked difference is the behavior of the many-body gap at large $U$. Unlike the pair-hopping model, where the many-body gap increases with $U$, the Hubbard model exhibits a non-monotonic dependence of the gap with $U$, with the maximum gap reached at $U\approx 0.4J$. The finite-temperature behavior of the SYK+Hubbard model is illustrated in Fig.~\ref{fig:Phase_diagram}, where we present the color plot of the logarithm of ODLRO on the temperature vs. $U/J$ plane. Once again, one notices the non-monotonic behavior of the critical temperature, where ODLRO is suppressed. 

The presence of the critical interaction strength, $U_c$, and the non-monotonic behavior of the gap and $T_c$ are contrary to the mean-field  predictions, Eq.~(\ref{eq:mean-field}). We attribute both phenomena to the strong quantum fluctuations in the SYK+Hubbard model.   
To account for such large $N$, non-mean-field phenomenology, we investigate below the SYK+Hubbard model in the two limiting cases of strong and weak attraction.  In both cases we are able to 
account for the quantum fluctuations and show that they indeed explain the observed behavior. 

In the case of the strong attraction 
this is achieved by  mapping onto an exactly solvable generalized Richardson model. It provides an asymptotically exact description of 
the low-energy part of the SYK+Hubbard model in the limit  $U\gtrsim \sqrt{N} J/13$. In the opposite limit of the weak attraction we reduce the problem to the quantum version of the Kuramoto model. It's classical counterpart \cite{Kuramoto1975,daido1992quasientrainment,wiesenfeld1998frequency,strogatz2000kuramoto,acebron2005kuramoto,arenas2006synchronization,gomez2007synchronizability,dorfler2013synchronization,boccaletti2014structure,witthaut2017classical,DSouza2019explosive} provides a paradigm for synchronization of non-linear oscillators. We show that the quantum Kuramoto model  provides description of the pseudogap phase for $U< U_c$ and the continuous superconducting QPT at $U=U_c$.


\section{Quantum fluctuations in SYK +Hubbard model}  
\label{sec:quantum-fluctuations}

\subsection{Generalized Richardson Model}
\label{sec:Richardson}

\begin{figure}[htb]
  \centering
  \includegraphics[width=0.5\textwidth]{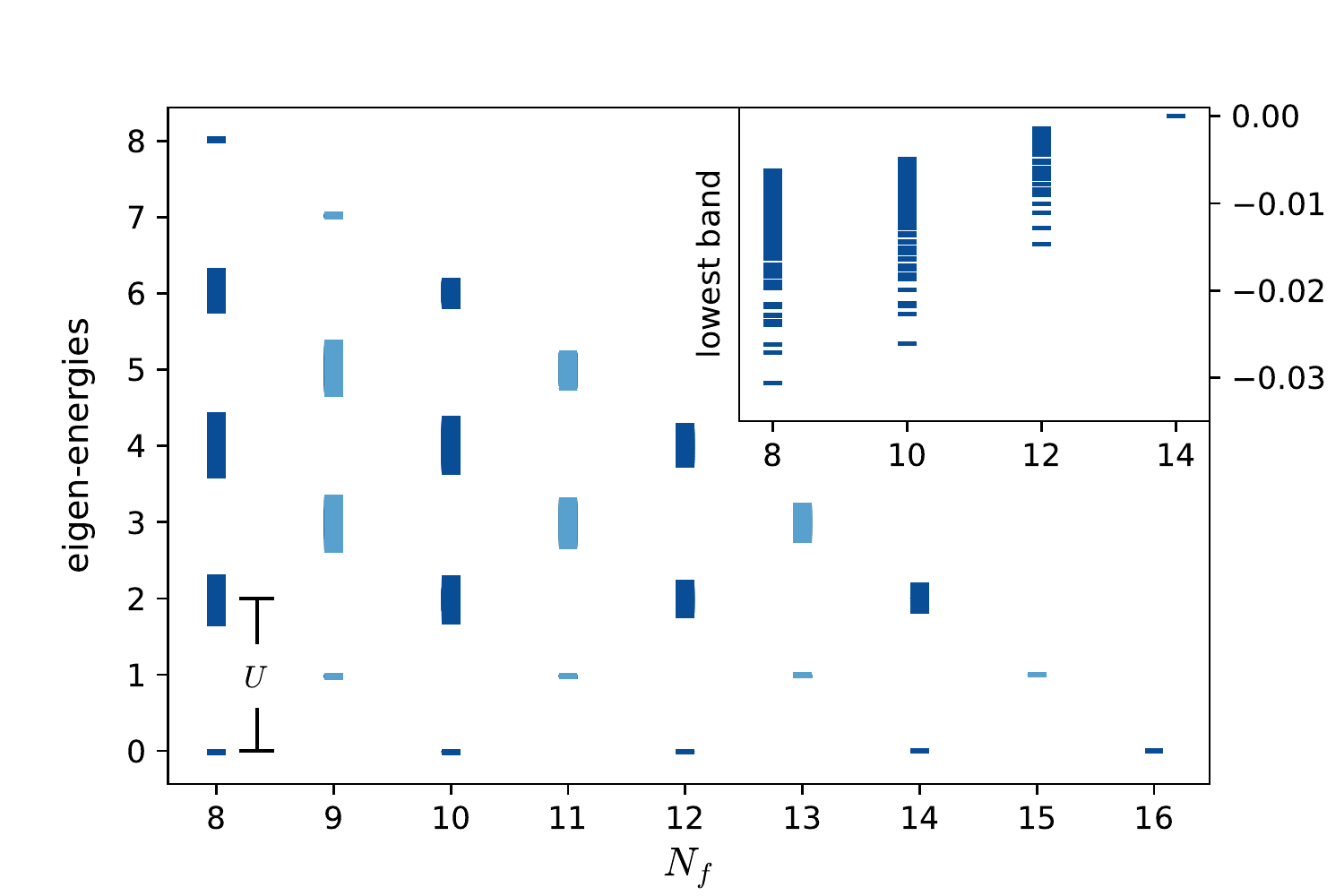}
  \caption{Many-body energy spectrum (in units of $J$) of SYK+Hubbard model vs. number of fermions $N_f$ for $N=8$ and $U=2J$. Chemical potential, $\mu$, is set  $\mu=-U/2$ to preserve particle-hole symmetry between $N_f$ and $2N-N_f$ sectors. The inset shows the lowest bands of the spectrum for even $N_f$. Band-width for the lowest energy sector at half-filling is consistent with $NW/16 \simeq 0.023$, predicted by the generalized Richardson solution, Eq.~(\ref{eq:spectrum}). }
  \label{fig:E_vs_Nf}
\end{figure}

The many-body spectrum of the SYK+Hubbard model with $U=2J$ and $N=8$ is shown in Fig.~\ref{fig:E_vs_Nf} as a function of the fermion number, $N_f$. 
One notices strong alternation of the ground state energies between even and odd fermion number. The low-energy part of the spectrum, which is not resolved in the main plot is shown in the inset for even $N_f$. These low-energy bands are separated by the gap $\sim U$ from the rest of the spectrum. Number of many-body states in these low-energy bands is exactly $\binom{N_f/2}{N}$, i.e. the number of ways to place $N_f/2$ indistinguishable pairs over $N$ orbitals. Therefore the low-energy bands are described by models of hard-core bosons, Eq.~(\ref{eq:bosons}). In the absence of the SYK term, bosons are localized and all $\binom{N_f/2}{N}$ bosonic states are degenerate with the energy $-U$ per boson. The SYK term induces an effective bosonic hopping and thus leads to a formation of the low-energy bands. 

To gain an insight in the physics of the corresponding bosonic model, consider a state with $N_f/2$ hard-core bosons occupying a subset of $N$ orbitals. 
Acting with a given term of the SYK Hamiltonian, (\ref{eq:model}), say $J_{ij;kl}$,  on such a state produces a non-zero result only if orbitals $k$ and $l$ are 
occupied, while $i$ and $j$ are empty (or vise versa). It leads to a state with $N_f/2-2$ bosons and 2 broken pairs (i.e. 4 unpaired fermions on orbitals $i,j,k,l$). Such a state costs energy $2U$ and resides outside of the low-energy bosonic sector. From the point of view of an effective bosonic model, it is a virtual state, which ought to be integrated out. To bring the system back to the bosonic sector one has to act on it with {\em the same} SYK term, $J_{ij;kl}$. 
This either brings the system back to the initial state (generating an uninteresting on-site energy shift), or results into hopping of {\em two} bosons from the orbitals $k,l$ to  $i,j$. The latter option gives rise to the effective bosonic Hamiltonian:
\begin{equation} \label{eq:bosonic-H}
  H_{\mathrm{b}}=-\frac{6}{2U} \sum_{ijkl}^N J_{ij;kl}^2\left[ b_i^{\dagger}b_j^{\dagger}b_k b_l + b_l^{\dagger}b_k^{\dagger}b_j b_i \right],  
  \end{equation}
where the factor of $6=2+4$ is coming from the opposite and same spin terms in the SYK Hamiltonian, correspondingly. There is also a one boson hopping term of the form $\sum_{jk} M_{jk}  b_j^{\dagger}b_k$, where $M_{jk}\propto - \sum_{il}^N J_{ij;kl}J_{lj;ki}/U$. Since the two matrix elements here are uncorrelated, the corresponding sum includes $N^2$ sign alternating terms, implying for a typical matrix element $|M_{ij}|\sim \sqrt{N^2}J^2/(N^3U)=J^2/(N^2U)$. This makes one boson hopping insignificant at large $N$.        

Hamiltonian (\ref{eq:bosonic-H}) represents a version of the bosonic SYK model \cite{fu2016numerical,Swingle2019}. Specifics of our model is that we 
work with {\em real} matrix elements $J_{ij;kl}$ and thus there is a {\em non-random} sign-definite part of the Hamiltonian (\ref{eq:bosonic-H}), which we call a generalized Richardson model:  
 \begin{eqnarray} \label{eq:Richardson}
   &&H_{\mathrm{gR}}\! =\! -\frac{W}{N^3}\! \sum_{ijkl}^N\! b_i^{\dagger}b_j^{\dagger}b_k b_l
  \\ && =\! -\frac{W}{N^3}\!\left[B_{0}^{\dagger} B_{0}^{\dagger} B_{0} B_{0} \!-\! 4 B_{0}^{\dagger}\hat N_b B_{0} + 2 \hat N_b(\hat N_b-1)\! \right]\!,
  \nonumber 
\end{eqnarray}
where $W= 3J^2/32U$ and all indexes $i,j,k,l$ must be distinct. We introduced operator $B_{0} = \sum_i^N b_i$ and the boson number operator $\hat N_b = \sum_i^N b_i^{\dagger}b_i$.  Employing the (anti)commutation relations for the hard-core bosons: $b_i^{\dagger}b_i+b_ib_i^{\dagger}=1$ and 
 $b_i^{\dagger}b_j - b_jb_i^{\dagger}=0$ for $i\neq j$,   one obtains
 \begin{equation}\label{eq:algebra}
  \begin{aligned}
  & [\hat N_b,B_{0}^{\dagger}] = B_0^{\dagger}; \quad [\hat N_b,B_{0}] =- B_0;
  \\ & [B_{0}^{\dagger},B_{0}] = 2\hat N_b-N. 
  \end{aligned}
\end{equation}
These operators form the su(2) algebra upon identification $\hat L_+ = B_0^{\dagger},\, \hat L_- = B_0,\, \hat L_z = \hat N_b-N/2$. 
One thus finds that: $B_{0}^{\dagger} B_{0}= \hat L^2 - \hat L_z^2 + \hat L_z$.  
This observation allows one to solve 
the Richardson Model \cite{Richardson1963,vonDelft2000,Dukelsky2004} with degenerate on-site energies, $H_{\mathrm{R}}=-\frac{W}{N}B_{0}^{\dagger} B_{0}$.  Let us focus for simplicity on the half-filled model, with $N_b=N/2$ and thus $L_z=0$. The spectrum of the half-filled Richardson model is thus  given by $E_\mathrm{R}(L)=-W L(L+1)/N$, where the total angular momentum runs $L=0,1,\ldots, N/2$. The unique ground state corresponds to $L=N/2$. The degeneracies of the excited states are given by the multiplicity of the corresponding representations: 
\begin{equation} \label{eq:degeneracy}
  \begin{aligned}
 D(L) = \binom{N/2-L}{N}-\binom{N/2-L-1}{N},
  \end{aligned}
\end{equation}
with the total number of states: $\sum\limits_{L=0}^{N/2-1}D(L)+1=\binom{N/2}{N}$, which is  the Hilbert space dimensionality for the half-filled hard-core particles.

In the same way one  finds the spectrum of the half-filled generalized Richardson model, Eq.~ (\ref{eq:Richardson}), to be:
\begin{equation} \label{eq:spectrum}
  \begin{aligned}
  & E_{\mathrm{gR}}(L) = -\frac{W}{N^3} [L(L+1)-(N-1)]^2 + \mathrm{const}, 
  \end{aligned}
\end{equation}
with the same set of degeneracies, Eq.~(\ref{eq:degeneracy}). The many-body gap between the ground state, $L=N/2$, and the first excited band with 
$L=N/2-1$ and degeneracy $D(N/2-1)=N-1$ is approaching  $W/2$ at large $N$.

The ground state is $|GS\rangle \propto (B_0^{\dagger})^{N/2} |0\rangle$. The corresponding single-particle density matrix $\rho_{ij}$, Eq.~(\ref{eq:density-matrix}),  has diagonal elements $\rho_{ii}=1/2$ and off-diagonal ones $\rho_{ij}=\frac{1}{4} \frac{N-2}{N-1}$. Thus its largest eigenvalue is
$\lambda_0 = N/4+1/2$ (dashed line in Fig.~\ref{fig:ODLRO}).  The fact that it scales as $N$ signals the presence of ODLRO in the ground state of the generalized Richardson model.  The remaining $N-1$ eigenvalues are degenerate at $\lambda_\alpha = \frac{1}{4} \frac{N}{N-1}$. These features are  qualitatively consistent with the exact diagonalization results of SYK+Hubbard shown in Fig.~\ref{fig:ODLRO} for  $U/J=2$.

\begin{figure}[htb]
  \centering
  \includegraphics[width=0.5\textwidth]{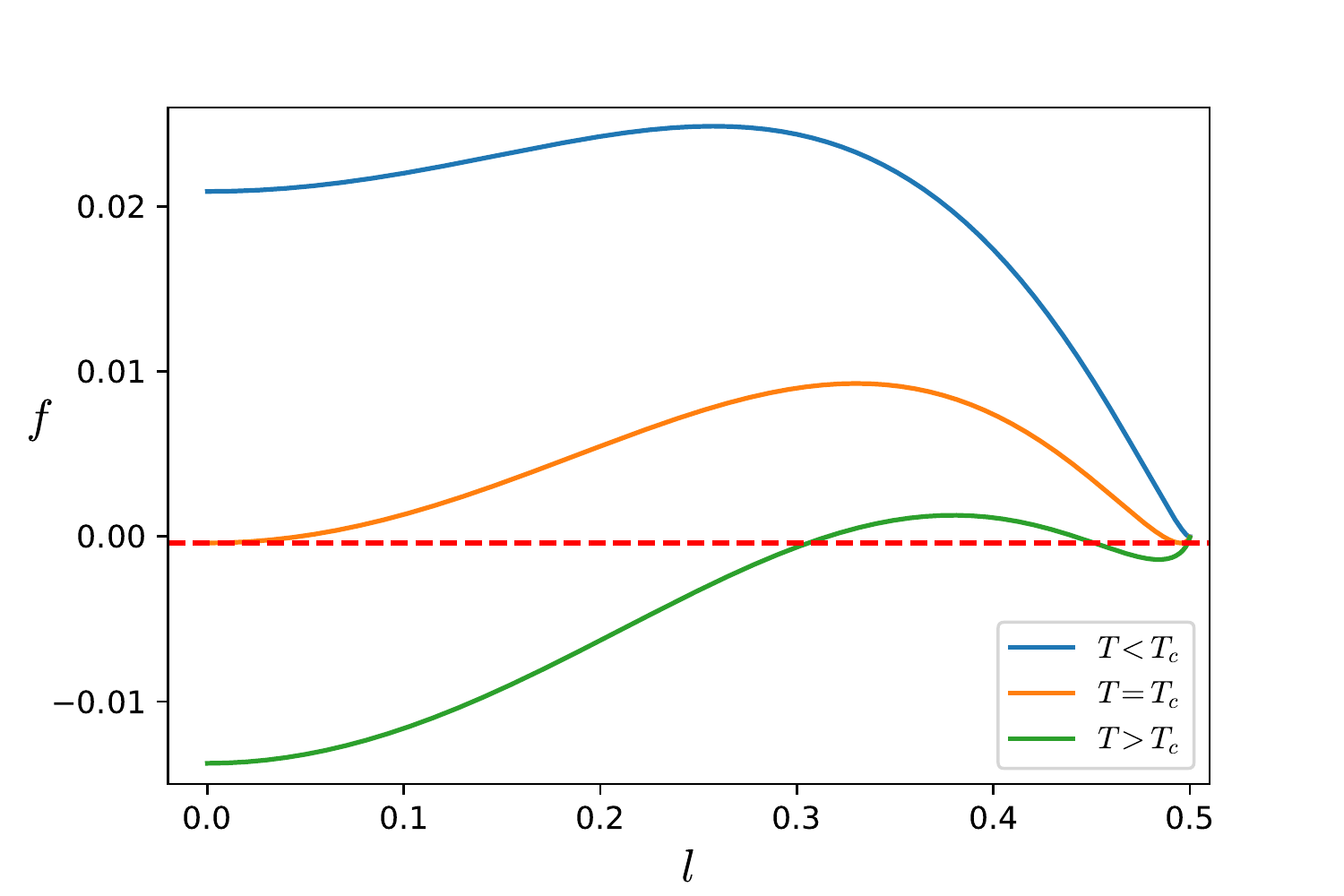}
  \caption{The free energy density of the generalized Richardson model, $f(l)$, Eq.~(\ref{eq:f}), vs. scaled ``angular momentum'' $l=L/N$ for different temperatures.}
  \label{fig:first_order}
\end{figure}

To access distraction of ODLRO at elevated temperature one considers the partition function: 
\begin{equation} \label{eq:partition}
  \begin{aligned}
   Z= \sum_{L=0}^{N/2} D(L) e^{- E_{\mathrm{gR}}(L)/T} \approx \int\limits_0^{1/2} dl\, e^{-N f(l)/T},
     \end{aligned}
\end{equation}
where we introduced $l=L/N$, substituting summation with the integration, and the free energy density, Fig. \ref{fig:first_order}, is defined as  
$f(l)=\lim_{N\to \infty}(E_{\mathrm{gR}}(l)-T\ln D(l))/N$:  
\begin{eqnarray} \label{eq:f}
   f(l) &=& -Wl^{\gamma}  \\
   &+& T \left[ (1/2-l) \ln (1/2-l) + (1/2+l) \ln (1/2+l)\right], \nonumber
\end{eqnarray}
where $\gamma=4$ for the generalized Richardson model.  

\begin{figure}[htb]
  \centering
  \includegraphics[width=0.5\textwidth]{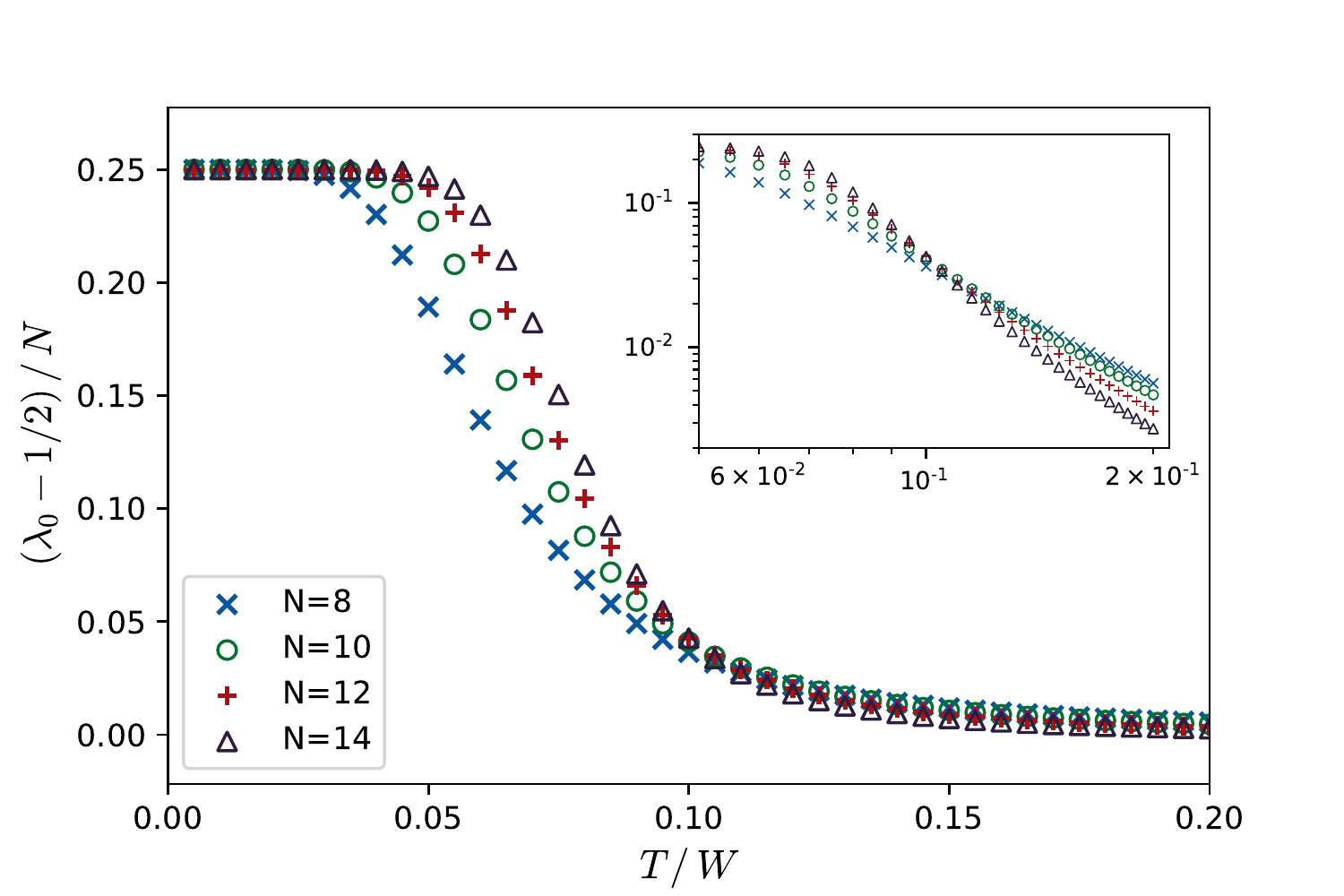}
  \caption{ODLRO vs. temperature for the generalized Richardson model, Eq.~(\ref{eq:Richardson}). Inset: vicinity of the crossing point. Compare with Fig.~\ref{fig:T-ODLRO} for the SYK+Hubbard model.}
  \label{fig:T-ODLRO_gR}
\end{figure}

In the large $N$ limit, the integral in Eq.~(\ref{eq:partition}) is dominated by the minima of $f(l)$. The latter 
changes from being $l=1/2$ at $T=0$ to $l=0$ at $T_c/W= 1/16\ln2\approx 0.09$, where the model undergoes the first order transition to a state with no ODLRO. This behavior is illustrated in Fig.~\ref{fig:T-ODLRO_gR}, which shows results of the exact diagonalization for the generalized bosonic Richardson model, Eq.~(\ref{eq:Richardson}). The crossing point at  $T/W \approx 0.1$ marks the first order transition, were ODLRO jumps from $1/4$ to zero in the $N\to \infty$ limit. This should be compared with the exact diagonalization of the SYK+Hubbard model shown in Fig.~\ref{fig:T-ODLRO}.    

It is instructive to compare this behavior with that of the traditional  
Richardson model, $H_{\mathrm{R}}=-\frac{W}{N}B_{0}^{\dagger} B_{0}$, whose partition function is again given by 
Eqs.~(\ref{eq:partition}), (\ref{eq:f}) with $\gamma=2$. The latter model may be seen to undergo a {\em continuous} phase transition 
at $T_c=W/2$. This model with $W=U$ is exactly the pure pair hopping model, Eq.~(\ref{eq:pair-hopping}).  

\begin{figure}[htb]
  \centering
  \includegraphics[width=0.45\textwidth]{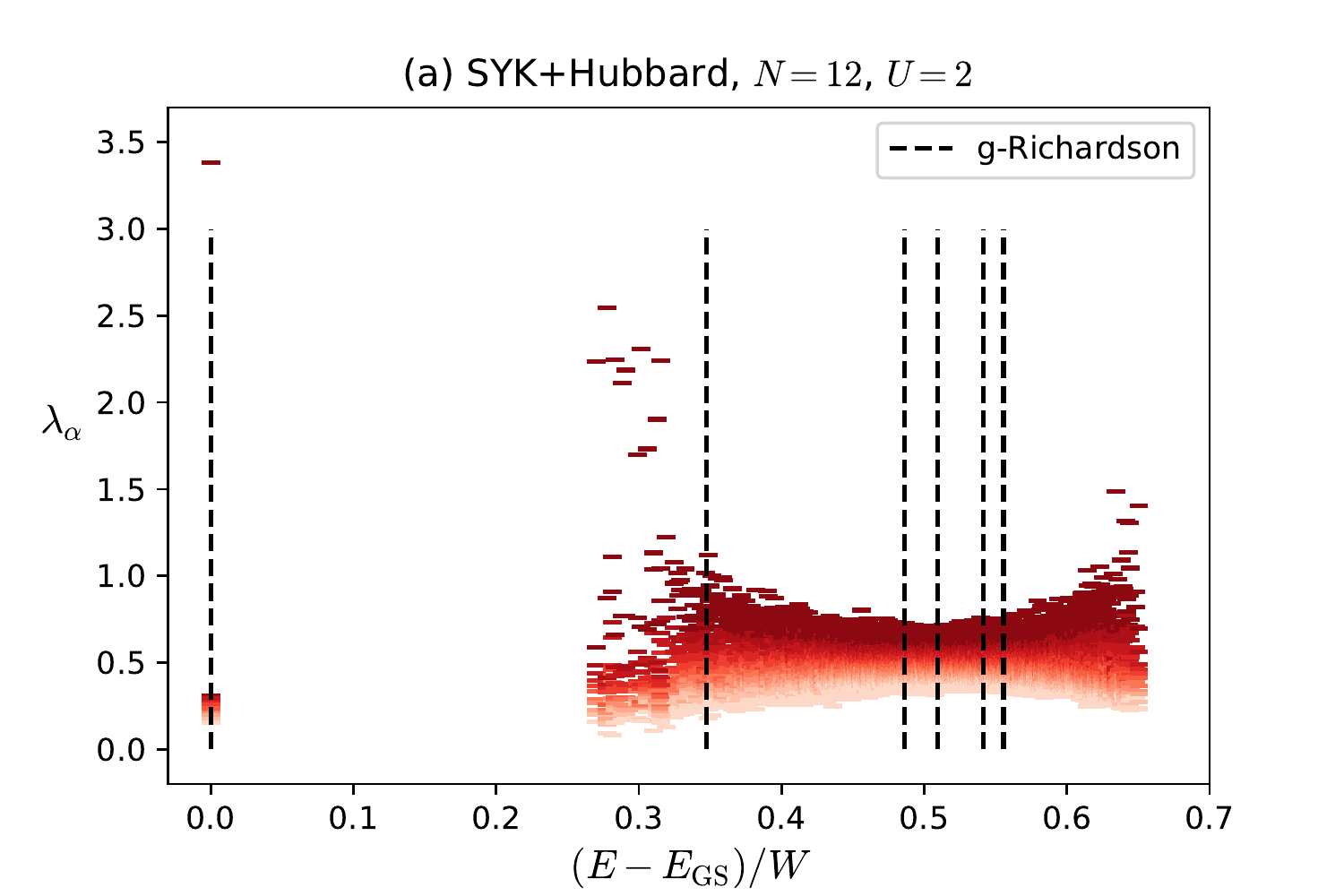}\\
  \includegraphics[width=0.45\textwidth]{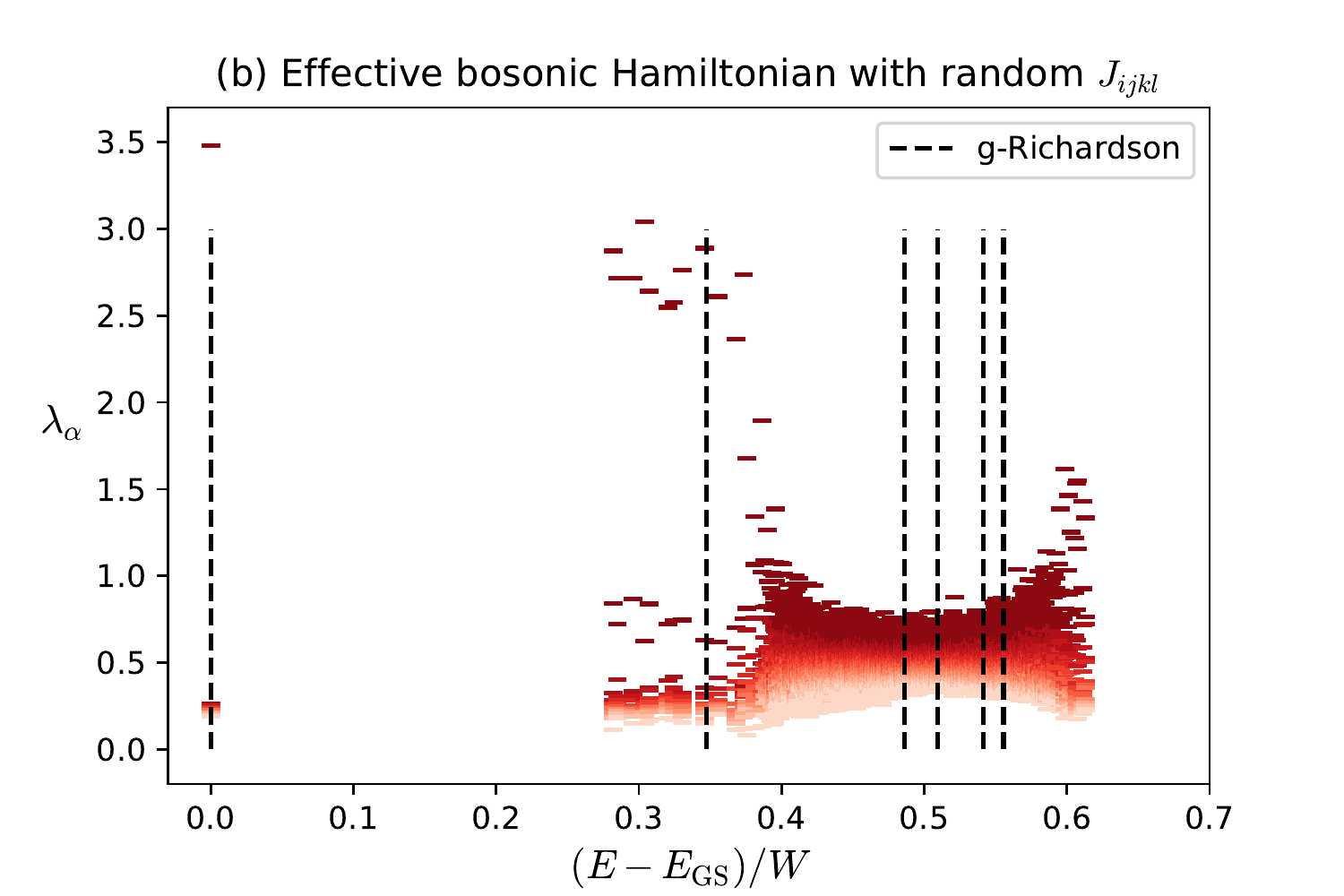}
  \caption{Spectra of $\langle n|b_i^{\dagger} b_j|n \rangle$ for each many-body state $|n\rangle$ vs. it's energy, $E-E_\mathrm{GS}$, for (a) SYK+Hubbard with $N=12$ and (b) effective low-energy bosonic theory with the Hamiltonian (\ref{eq:bosonic-H}). Black dashed lines are energies of the generalized Richardson model, Eq.~(\ref{eq:spectrum}).}
  \label{fig:spectrum-ODLRO_both}
\end{figure}

One may worry if the generalized Richardson model, Eq.~(\ref{eq:Richardson}), is a reasonable approximation for the low-energy 
bosonic model (\ref{eq:bosonic-H}). To answer this question one needs to examine the role of the random part of $J_{ij;kl}^2$ in Eq.~(\ref{eq:bosonic-H}).
This random part removes degeneracies, Eq.~(\ref{eq:degeneracy}),  between excited states with $L<N/2$, transforming them into the bands. Let's focus on the lowest 
such band with $L=N/2-1$, consisting of $N-1$ states. One can write an effective model for this band as $(N-1)\times (N-1)$ matrix 
Hamiltonian with the random elements $h_{rs}$.  Their variance can be estimated from the fact that  a matrix element $h_{rs}$ is given by a sum of 
$N^4$ random sign terms each of the order $W/N^3$. As a result, $\langle h_{rs}^2\rangle \sim N^4(W/N^3)^2=(W/N)^2$. The density of states 
of such random matrix is given by a semicircle with the bandwidth $\sqrt{N}W/N =W/\sqrt{N}$. Since the gap between the band and the ground-state scales as $W$, the latter remains well separated as long as $N\gg 1$ even for the random model, Eq.~(\ref{eq:bosonic-H}), see Fig.~\ref{fig:spectrum-ODLRO_both}.

We thus conclude that the generalized Richardson model, Eqs.~(\ref{eq:Richardson})-(\ref{eq:f}), provides an accurate description of the low-energy 
sector of the SYK+Hubbard model for $U\gg J$. It predicts ODLRO at low temperature. The many-body gap and critical temperature both scale
as $J^2/U$ with the large ratio between the two, $8 \ln 2 \approx 5.55$ (cf. with the BCS gap to $T_c$ ratio of $3.53$).  An enhancement of this ratio is also known in the context of quantum critical models  \cite{wu2019pairing},  holographic superconductors \cite{hartnoll2008building} and other SYK-like models \cite{patel2018coherent,esterlis2019cooper}.  These features are qualitatively consistent with the exact diagonalization results for the 
moderate $N$ SYK+Hubbard model. The single-particle fermionic excitations are separated by a larger gap $\sim U$. It is important to notice that the full bandwidth of the bosonic states is 
$NW/16= 3NJ^2/512 U$. The requirement for the Richardson model to be quantitatively accurate  is $U> 3 NJ^2/512 U$, i.e. $U \gtrsim \sqrt{N}J/13$.  This condition is satisfied for Figs.~\ref{fig:ODLRO}, \ref{fig:T-ODLRO} and \ref{fig:E_vs_Nf}.

\subsection{Pseudogap and the Quantum Kuramoto Model}
\label{sec:Kuramoto} 

We turn now to the opposite limit of $U\ll J$, where there is no separation between bosonic and fermionic sectors. To describe this limit, we notice that the action (\ref{eq:action}) exhibits a non-trivial saddle point with $|\Delta_i| = |\Delta| \propto Je^{-J\sqrt{\pi}/(8\sqrt{2} U)}$, Eq.~(\ref{eq:mean-field}). However, the phases, $\phi_i$, of the local order parameters,  
$\Delta_i =|\Delta|e^{i\phi_i}$,  are not fixed by the saddle point equations. They constitute thus the soft degrees of freedom, which are (almost) free to 
fluctuate. Such fluctuations are capable of destroying ODLRO, despite presence of the non-zero $|\Delta|$, even in the $N\to \infty$ limit.  

The action which governs the low-energy dynamics of the local phases is given by: 
\begin{equation}  \label{eq:Kuramoto}
  S[\phi_i(\tau)]\!=\!\!\int\! d\tau \left[ \frac{m}{2} \sum_i^N  \dot \phi_i^2  - \frac{g}{N} 
\sum_{i<j}^N\cos(\phi_i-\phi_j)
  \right]\!. 
\end{equation}
The two constants here, $m$ and $g$, are both related to the thermodynamic susceptibilities of the model. In principal, they are site specific, $m_i$ and $g_{ij}$, however in the large $N$ limit they may be substituted by the corresponding ensemble averages: $m=\overline{m_i}$ and $g=\overline{g_{ij}}$.   
The local compressibility $m_i=-\partial^2 E_\mathrm{GS}/\partial \mu_i^2$ is the susceptibility of the ground-state energy, $E_\mathrm{GS}$, to a local chemical potential, entering the Hamiltonian as $-\mu_i c^\dagger_{i\sigma}c_{i\sigma}$.  In the $|\Delta|=0$ case it was evaluated in Ref.~[\onlinecite{gu2019notes}] and found to be  $m\approx 1.04/J$. We do no expect it to be significantly affected by the presence of small $|\Delta|$. 
The off-diagonal Cooper susceptibility $g_{ij}/N=|\Delta|^2 \partial^2 E_\mathrm{GS}/\partial \bar\Delta_i \partial \Delta_j$ is a response to an extra term in the Hamiltonian  of the form $\bar \Delta_i c_{i\uparrow} c_{i\downarrow}+h.c.$. It is evaluated in Appendix \ref{app:B} and shown to be $g\sim |\Delta|^2/J$, with the mean-field pairing field $|\Delta|$ given by Eq.~(\ref{eq:mean-field}).

The action (\ref{eq:Kuramoto}) describes a quantum version of the celebrated classical Kuramoto model \cite{Kuramoto1975,daido1992quasientrainment,wiesenfeld1998frequency,strogatz2000kuramoto,acebron2005kuramoto,arenas2006synchronization,gomez2007synchronizability,dorfler2013synchronization,boccaletti2014structure,witthaut2017classical,DSouza2019explosive}. 
The latter was proposed \cite{Kuramoto1975} to describe  synchronization of coupled non-linear oscillators. It's quantum version, Eq.~(\ref{eq:Kuramoto}),  may be interpreted as  $N$-body quantum mechanics of particles with mass $m$ and coordinates $\phi_i$, residing on the unit circle and interacting via all-to-all $\cos$-potential. The synchronized phase of the classical Kuramoto model is analogous to a $\phi$-localized ground state  wavefunction of this quantum mechanics.  
Within the SYK+Hubbard model such synchronized phase means globally phase-coherent superconductivity with ODLRO. Below we show that the synchronized 
phase of the quantum Kuramoto model, Eq.~(\ref{eq:Kuramoto}), emerges above some critical coupling $g>g_c$ (i.e. at $U>U_c$) as a continuous 
QPT.   

Since the ground state is expected to be symmetric with respect to particle permutations, it may be thought off as a Bose condensate.
Due to all-to-all nature of the interactions, the Bose condensation in the large $N$ limit is accurately described by the  Gross-Pitaevskii equation. In the present context it takes the non-local form:
\begin{equation}  \label{eq:GP}
 -\frac{1}{2m}\frac{\partial^2 \Psi(\phi)}{\partial \phi^2} - \frac{g}{N} \! \int\limits_0^{2\pi}\! d\phi'\, |\Psi(\phi')|^2 \cos(\phi'-\phi) \Psi(\phi) = \mu \Psi(\phi),
\end{equation}
where the condensate wave-function is normalized as $\int_0^{2\pi}  \! d\phi\, |\Psi(\phi)|^2 =N$ and obeys the periodic boundary conditions, 
$\Psi(2\pi)=\Psi(0)$. Employing separability of the exponential 
potential, $e^{\pm i(\phi'-\phi)}$, one may reduce the non-linear equation (\ref{eq:GP}) to the linear Matheiu equation: 
\begin{equation}  \label{eq:Matheiu} 
 - \frac{1}{2m} \frac{\partial^2 \Psi(\phi)}{\partial \phi^2} - g \rho_1 \cos(\phi) \Psi(\phi) = \mu \Psi(\phi), 
\end{equation}
supplemented with the self-consistency condition 
\begin{equation}  \label{eq:rho1}
\rho_1 =\! \frac{1}{N} \int\limits_0^{2\pi}\! d\phi'\, |\Psi(\phi')|^2 \cos \phi', 
 \end{equation}
where $\rho_1$ the first Fourier harmonics of the normalized condensate density, $|\Psi(\phi')|^2/N$. The strategy is to find a ground state wave-function of the Matheiu equation (\ref{eq:Matheiu}) for a given amplitude of the $\cos$-potential, $g\rho_1$, and substitute it into the self-consistency condition (\ref{eq:rho1}) to find $\rho_1$.
A trivial solution, $\rho_1=0$, with the uniform condensate, $\Psi=\sqrt{N/2\pi}$, and $\mu=0$ exists for any $g$.  A non-trivial  solution with $\mu<0$ requires $g>g_c$. 

\begin{figure}[htb]
  \centering
  \includegraphics[width=0.5\textwidth]{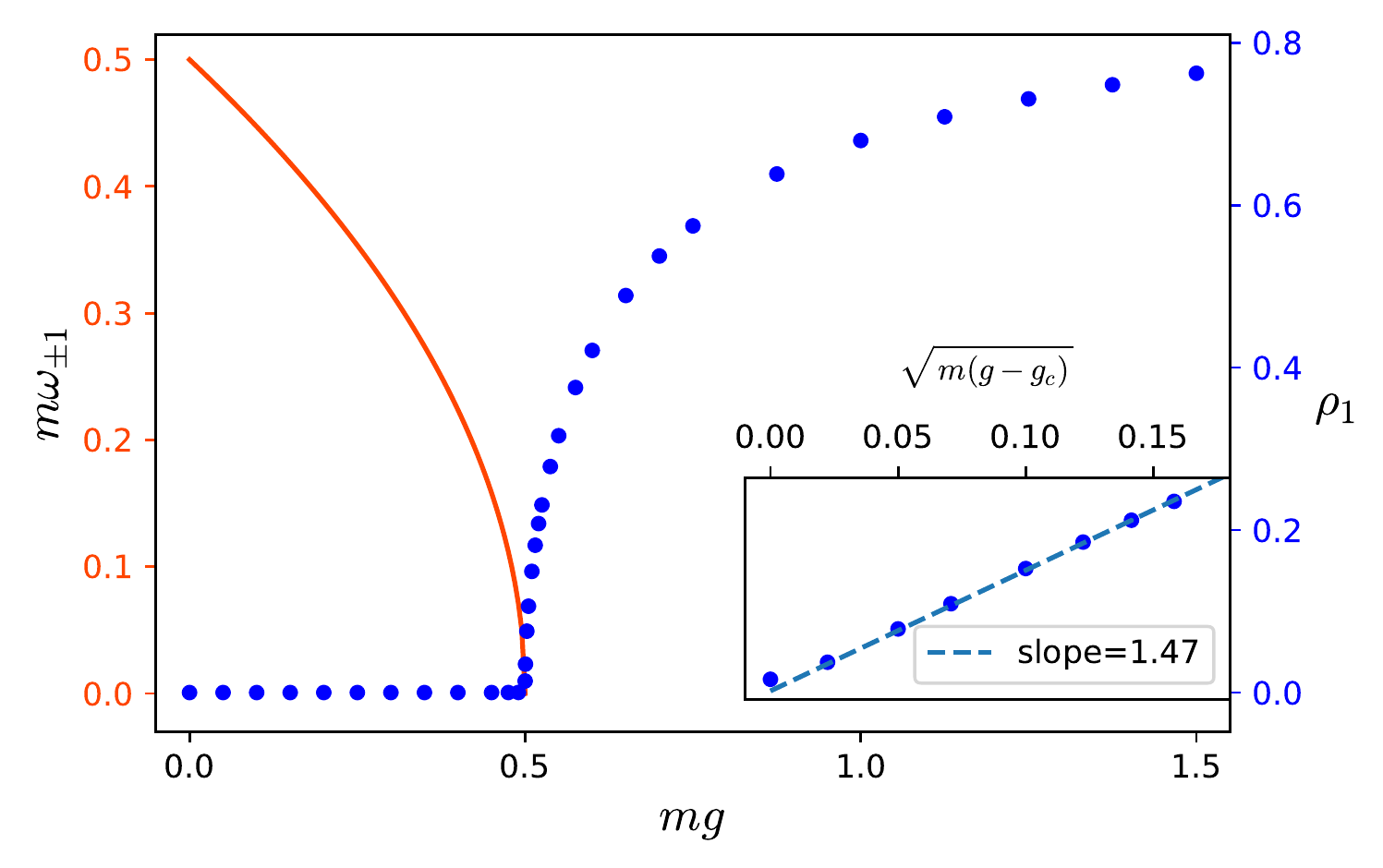}
  \caption{Numerical solution of Eqs.~(\ref{eq:Matheiu}) and (\ref{eq:rho1}) for the order parameter, $\rho_1$, of the quantum Kuramoto model (dots).
  The solid line is the frequency of the lowest Bogoliubov mode with $l=\pm 1$ for $mg<1/2$. Inset demonstrates $\rho_1\propto \sqrt{g-g_c}$ scaling for $g>g_c$.}
  \label{fig:rho_1-g}
\end{figure}

To find the non-trivial solution, one notices that the right hand side of Eq.~(\ref{eq:rho1}) is an odd function of  $g\rho_1$. Its behavior at small $g\rho_1$ may be found from the first order perturbation theory for the Matheiu equation (\ref{eq:Matheiu}), yielding the linear slope $2 mg\rho_1$. On the other hand, at large $mg\rho_1\gg1$ the ground state wave function of Eq.~(\ref{eq:Matheiu}) is a narrow Gaussian, centered at $\phi=0$. This implies that the right hand side of Eq.~(\ref{eq:rho1}) saturates to one for $mg\rho_1\gg 1$. As a result, Eq.~(\ref{eq:rho1}) is the standard mean-field equation for a second order transition with the order parameter $\rho_1$. It yields a finite order parameter $\rho_1 \propto \sqrt{g-g_c}$ for $g\gtrsim g_c$ with $g_c=1/2m\approx 0.48 J$, Fig.~\ref{fig:rho_1-g}.

An alternative way to determine $g_c$ is to investigate a spectrum of linearized fluctuations on top of the uniform solution,  $\Psi(\phi,t)=\sqrt{N/2\pi} + 
\sum_l \psi_l e^{il\phi -i\omega_l t}$, where $l=\pm 1,\pm 2,\ldots$ labels angular momentum components. Substituting this into the time-dependent Gross-Pitaevskii equation, Eq.~(\ref{eq:GP}) with $i\partial_t \Psi$ on the right hand side, and linearizing it with respect to $\psi_l$, one finds the spectrum:
\begin{equation}   \label{eq:normal-modes}
\omega_{\pm 1}=\sqrt{\left(\frac{1}{2m}\right)^2-\frac{g}{2m}};\quad\quad  \omega_{|l|\geq 2} =\frac{l^2}{2m}. 
\end{equation}
Therefore for $g>g_c=1/2m$ the frequency of the $l=\pm 1$ components becomes imaginary, indicating instability towards a non-uniform condensate. 
This expression shows that the continuous QPT is indeed associated with the time-scale $\omega_{\pm 1}^{-1}\propto |g_c-g|^{-z\nu}$, which is divergent at the transition with the  Gaussian exponent $z\nu =1/2$.    

We thus conclude that the quantum Kuramoto model exhibits the synchronized phase for $mg> 1/2$, where the local phases, $\phi_i$, are coallesing. 
In the large $N$ limit this spells spontaneous breaking of the $U(1)$ symmetry. In terms of the SYK+Hubbard model these observations translate 
into formation of ODLRO for $U>U_c$, where, employing Eqs.~(\ref{eq:mean-field}) and (\ref{g_Kuramoto}), $U_c\approx  J\sqrt{\pi}/(4\sqrt{2}\log C_2)$, see Appendix \ref{app:B}.   The quantum Kuramoto model synchronization transition is indeed  seen in the exact exact diagonalization of the SYK+Hubbard model, Fig.~\ref{fig:Hubbard-U}, as the continuous QPT at  $U=U_c$.

For $U<U_c$ the on-sites phases 
$\phi_i$ fluctuate freely and prevent formation of the global ODLRO. This phenomenon renders the mean-field treatment of Sec.~\ref{sec:mean-field} 
grossly inadequate for $U<U_c$ and leads to creation of the {\em pseudogap} phase. The latter is characterized by the even-odd alternation in the 
ground state energies, cf. Fig.~\ref{fig:E_vs_Nf}, thus exhibiting a {\em single-particle} energy gap (i.e. a finite energy to add or subtract a single fermion). However, there is neither ODLRO nor the  many-body gap within a sector with a fixed $N_f$.  Therefore from the transport perspective, the pseudogap state is characterized as an insulator. Correspondingly the Kuramoto QPT should be termed  an insulator--superconductor one.  

The line $2\pi T\approx \omega_{\pm 1}$, Eq.~(\ref{eq:normal-modes}), spells the boundary of the quantum critical regime.
If $2\pi T<\omega_{\pm 1}$, the quantum Kuramoto phase fluctuations, governed by $\langle e^{i\phi_i(\tau)} e^{-i\phi_i(0)}\rangle= e^{-\omega_{\pm 1}|\tau|}$, are averaged out to zero. This  leads thus to the familiar SYK non Fermi liquid fermionic correlations. However, for 
$\omega_{\pm 1}<2\pi T\lesssim |\Delta|$ the imaginary time circle is too short to completely wash out the superconducting correlations. This creates an interesting quantum critical scenario, where superconducting correlations show up as a finite temperature effect.        

%

\section{Towards a holographic interpretation}
\label{sec:holography} 

In this section we  briefly comment on a possible holographic interpretation 
of our findings. Recall that at $T=0$ we have seen formation of local  Cooper pairs 
at  arbitrary small attraction between fermions.  Their phases are incoherent 
at intermediate $U$,  separated by the continuous QPT from  the superfluid phase with ODLRO 
at large $U$.   The complex SYK dot we work with is now used as a toy model for ``near AdS/almost CFT"
correspondence in quantum mechanics. From a higher-dimensional perspective
the Reissner-Nordstrom (RN) black hole (BH) is considered as the bulk whose geometry involves a long AdS$_2$ 
throat near the horizon. The large $N$ SYK quantum mechanics lives at the boundary of the throat and conjecturally
is dual to the AdS$_2$ near horizon, flavored with some matter.

The effective low energy boundary action for the unperturbed complex SYK involves two Goldstone modes - the reparameterization of time, governed by the Schwarzian action, and the $U(1)$ phase field, $\phi(\tau)$, governed by the kinetic term $\dot\phi^2$ (see, for example \cite{gu2019notes} and references therein). The bulk theory in addition to Jackiw-Teitelboim (JT) gravity involves  the  gauge fields and some matter. If we focus at $T=0$ the extremal RN BH  is unstable under  small perturbations. The mode of instability can be interpreted as the Schwinger pair creation (see \cite{hartnoll} for a review).
Usually the bulk instability is treated as formation of the homogeneous condensate 
described in terms of the boundary behavior of the bulk complex scalar or the bulk fermion.

The individual local Cooper pairs play an important role in our analysis hence
their holographic meaning needs to be clarified. Let us start with the bulk identification of the 
Goldstone phase field. To this aim consider for a moment the $U(1)$ bulk $2d$ field $(A_\tau, A_r)$
with the boundary behavior involving chemical potential and  density
\begin{equation} 
A_{\tau} (r\rightarrow 0)= \mu + \rho r, \qquad F_{\tau, r}=\rho.
\end{equation} 
In the boundary theory the density, $\rho$, and the phase, $\phi$,  are conjugated variables
\begin{equation}
[\rho,\phi]=i.
\end{equation}  
Hence the phase has to be canonically conjugated to $F_{\tau r}$ in the bulk.  To get
the correct conjugated variable recall 
the canonical pair in $2d$ gauge theory
\begin{equation} 
[E_r(\tau,r),A_r(\tau,r')]=i\delta (r-r'), 
\end{equation}
which allows to identify the phase field, $\phi(\tau)$, as the gauge holonomy along the radial direction, $r$.
\begin{equation} 
\phi(\tau)=\int\! dr\, A_r(\tau, r). 
\end{equation}
Note that if we choose $A_r=0$ gauge, the holonomy factor appears in the boundary conditions.

\begin{figure}[htb]
  \centering
  \includegraphics[width=0.4\textwidth]{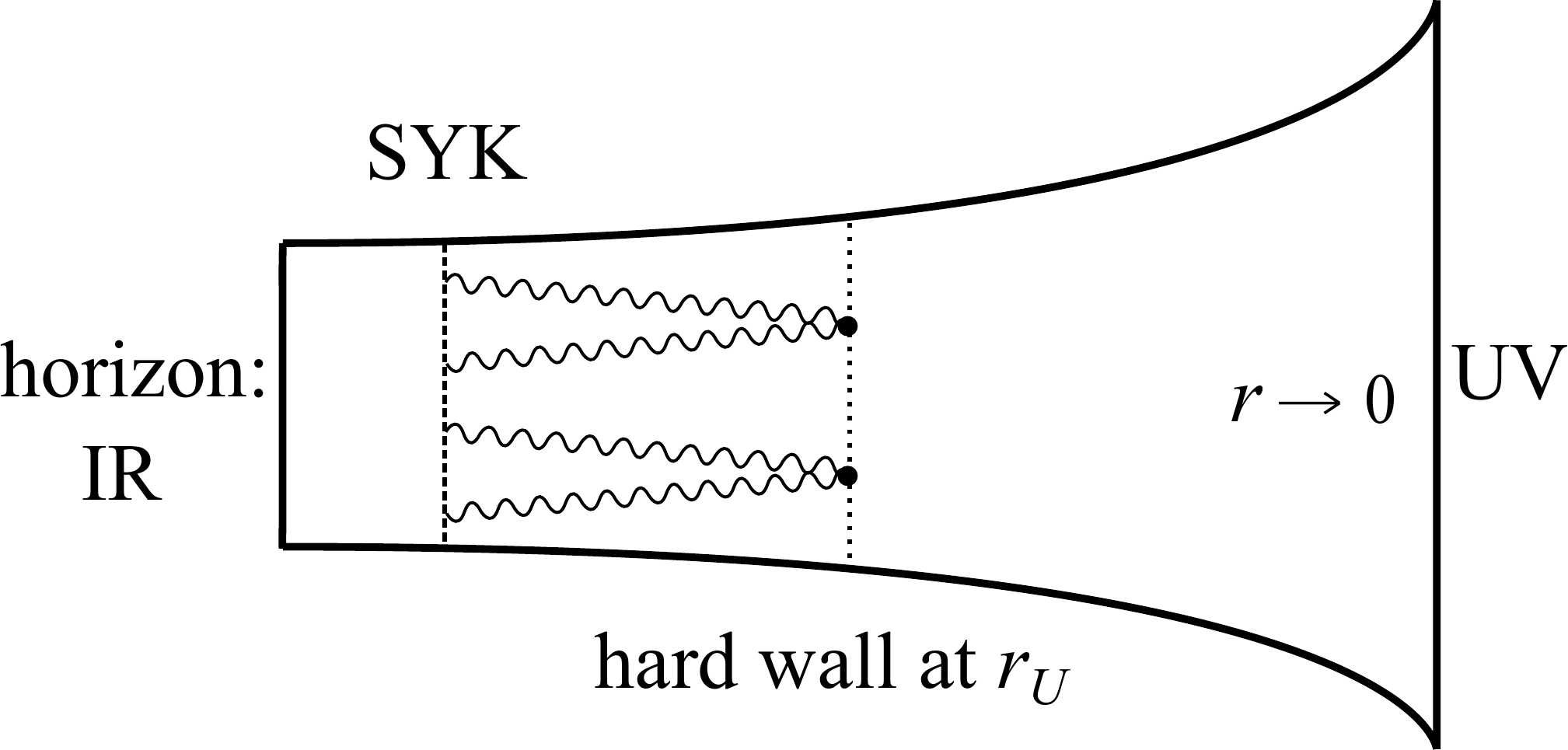}
  \caption{ Schematic picture of the conjectured bulk representation of Cooper pairs.
The SYK and the hard wall are placed in the throat region of the BH geometry.
A baryonic vertex is near the radial position of hard wall and gets
fractionalized into constituents supporting two strings each. } 
  \label{fig:holography}
\end{figure}

A similar identification of the Goldstone phase modes has been 
developed in holographic QCD \cite{ss} and in the holographic hydrodynamics
\cite{son2}. In QCD the bulk flavor gauge group $SU(N_f)_L\times SU(N_f)_R$
is broken by the Higgs mechanism down to the diagonal $SU(N_f)$ and the 
pions $\pi^a$, which are non-abelian Goldstone phases, of the chiral (excitonic) condensate 
are identified as $\exp(i\pi^a t^a) = P \exp \int dr A_r(r,x)$. In the
holographic hydrodynamics  a similar identification of the Goldstone phase is 
emerging upon breaking of $U(1)\times U(1)$ symmetry to the diagonal $U(1)$  \cite{son2}.

We turn now to the interpretation of the Hubbard  $U$.
Fortunately, the Hubbard model has been treated in the holographic approach for 
Bose \cite{meyer1} and Fermi systems \cite{meyer2}, where it was realized that  the 
Hubbard $U$ is to be identified with the radial position of the hard wall
$r_U=U$. Therefore the control parameter, $U/J$,  
tells how close to the horizon the hard wall
is placed. Small $U$ corresponds to  IR near horizon region,  while 
large $U$ corresponds to the hard-wall at UV near the boundary of AdS$_2$. 
As follows from our analysis,  a perturbation induced by the IR wall at small $U$ amounts to the 
instability of the extremal RN geometry and formation of the Cooper pairs. 
At large $U$ the gap of an individual Cooper pair, $\Delta \sim U$, fits
the  length of two strings extended up to the $U$ scale, representing two fermions at
the boundary.

The most subtle question concerns the  identification of the bulk 
counterpart of  Cooper pairs. To formulate the conjecture let us remind  
universal aspects of  the instanton   solutions in different dimensions. 
Consider, for instance, instanton
in the $SU(N)$ gauge theory on the $R^3 \times S^1$ geometry. If there is 
non-vanishing holonomy around $S^1$ the instanton with unit topological charge
is split into $N$ constituents with the topological charges $1/N$ and
non-vanishing monopole charges \cite{caloron} (the total monopole charge is zero, of course). 
This is the ``caloron''
solution, known for the theory with finite temperature, or with  one compact 
space coordinate. The positions of the 
constituents $(x_1,\dots ,x_N)$ are fixed by the eigenvalues of the 
$SU(N)$ holonomy, $V$, around $S^1$, $V= \mathrm{diag}(e^{ix_1}\dots ,e^{ix_N})$. 

The holographic Skyrmion solution in $D=4$ is the instanton in the gauge theory with the flavor
gauge group \cite{son}, which involves three space coordinates and the radial
coordinate $A_{\mathrm{inst}}(x_1,x_2,x_3,r)$. In the string theory framework equivalently it
is  a baryon vertex identified with the particular D-brane wrapped around the 
internal sphere \cite{witten}. Due to the anomaly,  $N$ strings are attached to the baryon vertex
located at some point in the bulk which amounts to $N$ fermions at the boundary. 
In holographic QCD the baryonic vertex is placed dynamically nearby the
effective IR wall \cite{ss}, which at small $U$ corresponds exactly to our $r_U=U$ scale.

Let's assume for a moment that our large $N$ SYK+Hubbard dot is a kind of Skyrmion-instanton
state that is a baryon vertex placed at $r_U$ in the throat region like in
holographic QCD. $N$ strings are attached to the vertex hence this picture at the first glance has
nothing to do with the ensemble of Cooper pairs. However, we have identified 
the  Goldstone phases as the holonomies in the radial coordinate and, as was established here, 
these holonomies do not vanish. Since the Skyrmion-instanton
solution involves the radial coordinate,   the nontrivial $SU(N)$ radial holonomy
splits the Skyrmion-instanton into Skyrmion-caloron with  at most $N/2$ 
fractional Skyrmion constituents with a fractional ``topological charge". Such
fractional Skyrmions  host now {\em two} strings instead
of $N$ strings and therefore amount to the pair of fermions at 
the boundary. Hence the fractional Skyrmion is a candidate for a
bulk counterpart of an individual  Cooper pair. Similarly to the standard caloron, the 
holonomies of the Cooper pairs correspond to the positions of constituents
on the ``dual circle" providing the Kuramoto-like picture for the individual
phases.

Establishing a potential for phases or the fractional Skyrmions  is a dynamical issue.
The potential for the phases of constituents of 
the caloron solution involves perturbative and 
non-perturbative contributions  and typically reads as 
$V(\phi_i)=\delta \sum_{i,j} \cos(\phi_i - \phi_{i-1})$ where
$\delta$ is a gap in the model.
I.e. typically potential for the phases 
involves only nearest neighbors interaction. In the case of Skyrmion-caloron
the all-to-all SYK Hamiltonian apparently induces an all-to-all interaction between
phases of individual components.

Summarizing, we conjecture that there is the baryonic vertex placed 
nearby $r_U$ radial scale. The non-trivial phases, corresponding to the
radial holonomies $\exp(i\phi_i)$ of the $U(1)_i$ bulk gauge field from the total $U(1)^N$
flavor gauge group, result in the splitting of the Skyrmion supporting
$N$ strings into the fractional Skyrmions supporting {\em two} strings. 
The disorder SYK Hamiltonian induces the all-to-all Kuramoto potential
for the phases of the Cooper pairs and the phases become 
synchronized at some position of the hard wall specified by $U_c$.
Of course,  many aspects of this conjecture deserves clarification.

Note some analogy with QCD at non-vanishing density. It is well-known
that at large baryonic density QCD is in the color-flavor locking
phase with the Cooper condensate of quarks. However it was argued in
\cite{rho} that at smaller chemical potential there is a transition
from Skyrmions into half-Skyrmions. It is assumed that at the transition
the common gap and exciton(chiral) condensate disappears.   Still there
are ``islands'' of gapped phase with disordered chiral phases. This resembles
the behavior of our model near the QPT.
  
Two additional remarks are in order. The insulator-superfluid QPT 
in $2+1$ has been discussed in the holographic framework in \cite{takayanagi}
and has clear parallels with our $0+1$ case. The insulator phase
was related with the AdS soliton background while the superfluid phase
with the AdS BH background. The AdS soliton solution has the effective
IR cut-off at a tip of the cigar, which is an analogue of our small $U$ regime,
since $U$ provides the IR cut-off as well. When $U$ is large it no longer
serves as  an IR parameter, yielding the UV scale instead. The BH physics
starts to  dominate in the superfluid phase in IR similar to our case.   

Another point concerns the origin of the Hubbard perturbation of the
SYK model. We have chosen it by hand, but it has appeared in the 
holographic setup in an interesting manner in large $N$  $\cal{N}$=4 SYM theory \cite{serban}. 
Namely, consider 
a string moving in $S^3$ or equivalently study the anomalous 
dimensions of the  particular scalar operators with large 
conformal dimension in the boundary theory. In this scenario,  
the dilatation operator at three loop exactly  coincides with the Hubbard Hamiltonian.
The latter plays the role of a conventional Hamiltonian of the discretized string, propagating in the nontrivial background.
Parameter $U$ in the Hubbard model is identified with an inverse coupling
in the boundary theory. It is unclear at the moment if these ideas provide an additional
intuition for our model.
The discussion in this section is clearly only qualitative and tentative. We postpone a more detailed
analysis of the  holographic picture for a separate study.

\section{Conclusion and Outlook}   
\label{sec:conclusions} 

Following the earlier studies \cite{esterlis2019cooper,*hauck2019eliashberg,wang2019solvable,patel2018coherent,chowdhury2019intrinsic,*chowdhury2019unreasonable}, we found that the spin-full version of the SYK model with an extra attractive interactions may exhibit ODLRO and superconductivity. Furthermore we found that details of this extra attraction are crucially important in dictating the global phase diagram of the model. The previous  studies focused on an effective all-to-all attraction, which conform to the large $N$ mean-field treatment. The latter calls for superconducting instability of the non Fermi liquid groundstate at an arbitrarily weak attraction. This is indeed the case for the SYK+pair hopping model briefly considered here.

Our main finding is that a local attraction, such as on-site negative $U$ Hubbard term, leads to a qualitatively different scenario of the superconducting transition. In this case the physics is dictated by quantum fluctuations of local phases. They destroy ODLRO in a sizable part of the phase diagram, confining 
the superconductivity to a dome-like region, Fig.~\ref{fig:Phase_diagram}. In particular, they lead to the pseudogap phase at small $U$ and the continuous QPT to the superconducting phase at $U=U_c$. These features are described by the quantum version of the celebrated Kuramoto model. At strong attraction, the local nature of the attractive interactions is also of crucial importance, resulting in $T_c\sim U^{-1}$ scaling of the critical temperature. This limit is mapped on the Richardson-like model with two-boson hopping.  It's exact solution predicts the first order transition at $T=T_c$ from ODLRO into a bosonic insulator state. The latter consists of fermions, paired with the binding energy $U\gg T_c$, forming a gas incoherent bosons. Fermion transport in this state is suppressed as $e^{-U/T}$.     

We list now some of the open  questions raised by our study: (i) What are fermionic correlation functions in the pseudogap phase at $U<U_c$? The naive 
answer is that they are the same as in the non Fermi liquid SYK model. Yet, contrary to SYK, fermions interact with the dynamical  phases as 
$|\Delta|e^{i\phi_i(\tau)}c_{i\downarrow}c_{i\uparrow} +h.c.$, where the phases, $\phi_i(\tau)$, are governed by the Kuramoto quantum mechanics, Eq.~(\ref{eq:Kuramoto}). Close to the QPT this dynamics becomes increasingly slow, Eq.~(\ref{eq:normal-modes}), and may significantly alter the fermionic 
correlation functions.    
    
(ii) What are the implications of our 0D treatment for the array geometry? In particular, is the dome-like phase diagram, Fig.~\ref{fig:Phase_diagram}, applicable to arrays and how it depends on the coupling (hopping) strength between the dots in the array?   

(iii) Is there an interaction and an interplay between the phases, governed by the Kuramoto, and the reparametrization modes \cite{Kitaev2015,Bagrets-Altland-Kamenev2016}, governed by the Schwarzian action? The latter modes are described by the Liouville quantum mechanics \cite{Bagrets-Altland-Kamenev2016}, which predicts metal-insulator crossover at the energy scale $J/N$. For a finite $N$ this energy scale may compete with the many-body 
gap $|\Delta|^2/J$, possibly affecting the insulator-superconductor QPT \cite{altland2019quantum}.    

 
(iv) An interesting generalization is a model with a weak time reversal symmetry  breaking
parameter. In the Richardson model such generalization leads to the Russian Doll (RD) model, Appendix \ref{app:Richardson}, 
which is known to be integrable.
One may expect that deformed in this manner the large $U$ generalized Richardson 
is also integrable. SYK corresponds to the completely degenerate local
Richardson parameters, $\epsilon_i=0$, which means that holographically all  
flavor branes are sitting on the top of each other in the IR and the $SU(N)$ symmetry is classically
unbroken. Generic values of $\epsilon_i$  correspond to displacements  
of flavor branes in the radial coordinate in the holographic treatment of 
Richardson or RD models. It would be interesting to elucidate 
the role of non-vanishing local parameters, $\epsilon_i$, in the generalized Richardson model.

(v) The quantum Kuramoto mechanism of the condensate formation could fit within a 
more general framework. In particular, an intermediate  pseudogap phase is believed 
to exist in the thermal QCD below the deconfinement phase transition, where the local phases
of the  chiral condensate are disordered. The synchronization of the chiral phases leading
to formation of the homogeneous chiral condensate may occur in a  Kuramoto-like way. 
Indeed as shown above, at the $1/N$ order the near-horizon
gravity (RG) dynamics induces the Kuramoto potential for  phases of the local Cooper pairs.
Formation of the chiral condensate in the holographic QCD, being also a near-horizon effect,  
may thus lead  to a non-abelian generalization of the Kuramoto potential for the exciton pairs.

\begin{acknowledgments}
We are grateful to A. Chubukov, A. Klein and J. Schmalian  for useful discussions. H.W. and A.K. were supported by  NSF grant DMR-1608238.
A.C. and A.G. thank the Fine Theoretical Physics Institute  at the University of Minnesota for hospitality and support. The work of A.G. was supported
by Basis Foundation fellowship and RFBR grant 19-02-00214.

\end{acknowledgments}

\appendix 

\section{Mean-field treatment of SYK-Hubbard model}
\label{app:mean-field}
In this Appendix we provide details of the mean-field treatment for the model specified by Eqs. (\ref{eq:model})--(\ref{eq:Hubbard}). We employ the standard treatment of SYK model, which  includes averaging of the replicated partition function over the distribution of couplings followed by the so-called $G \Sigma$-approach \cite{Bagrets-Altland-Kamenev2016}. For the model with {\em real} couplings, Eq.~(\ref{eq:model}), the Gaussian averaging over $J_{ij;kl}$'s produces two kinds of  8-fermion terms, which we call normal and anomalous 
\begin{eqnarray}
 && 
\left\langle e^{-\sum_{a=1}^n \int H_a d\tau}\right\rangle_{J} = \\
\nonumber && 
\exp\left\{ \frac{ J^2}{4(4 N)^3}\!\! \sum_{a,b=1}^n\! \int\! \!d\tau d\tau' \!\!\!
\sum_{i,j,k,l=1}^N\!\!\! \left(\mathcal{A}_{ijkl}^{a\tau,b\tau'} \! \!+ 
\mathcal{N}_{ijkl}^{a\tau,b\tau'}  \right) \! \right\},
\end{eqnarray} 
where the anomalous part $\mathcal{A}_{ijkl}^{a\tau,b\tau'}$ is given by a product of fermion operators describing creation and  annihilation of on-site Cooper pairs 
\begin{equation}    \label{eq:A-anomalous} 
\mathcal{A}_{ijkl}^{a\tau,b\tau'}=\sum_{\sigma\sigma'\rho\rho'} (\bar{c}_{i\sigma}^{a\tau}\bar{c}_{i\rho}^{b\tau'}) (\bar{c}_{j\sigma'}^{a\tau}\bar{c}_{j\rho'}^{b\tau'}) (c_{k\sigma'}^{a\tau}c_{k\rho'}^{b\tau'}) (c_{l\sigma}^{a\tau}c_{l\rho}^{b\tau'})
\end{equation}
and the normal part is given by product of one creation and one annihilation operator at each site 
\begin{equation}
\mathcal{N}_{ijkl}^{a\tau,b\tau'}=\sum_{\sigma\sigma'\rho\rho'} (\bar{c}_{i\sigma}^{a\tau}c_{i\rho}^{b\tau'}) (\bar{c}_{j\sigma'}^{a\tau}c_{j\rho'}^{b\tau'}) ( \bar{c}_{k\rho'}^{b\tau'}c_{k\sigma'}^{a\tau}) (\bar{c}_{l\rho}^{b\tau'}c_{l\sigma}^{a\tau}).
\end{equation}
Guided by the knowledge that no replica-off-diagonal saddle points exist for the SYK-model \cite{CommentsSYK16,Bagrets-Altland-Kamenev2016,wang2019replica}, we restrict further consideration to the replica-diagonal sector and drop the replica indexes hereafter. 
In the framework of  $G \Sigma$-approach one  introduces fields corresponding to the on-site  Green's functions. However, the presence of the anomalous term, Eq.~(\ref{eq:A-anomalous}), requires introduction of both normal and anomalous Green's functions. 
Anticipating spin-singlet superconducting pairing, we assume the anomalous fields $\bar{F}, F$ to have nonzero componens for the opposite spin-indexes only, such as 
\begin{equation}
F_{\tau\tau'}=-\frac{1}{N}\sum_{i=1}^N c_{i\downarrow}^{\tau} c_{i\uparrow}^{\tau'}, \, \, \,  
\bar{F}_{\tau\tau'}=- \frac{1}{N}\sum_{i=1}^N\bar{c}_{i\uparrow}^{\tau} \bar{c}_{i\downarrow}^{\tau'}.
\end{equation}
In contrast, since we do not expect magnetic ordering, the normal fields $G$ and $\Sigma$ are assumed to have nonzero components only for the coinciding spin-indexes, 
\begin{eqnarray}
\nonumber && 
G_{\tau\tau'}=-\frac{1}{N}\sum_{i=1}^N \bar{c}_{i\sigma}^{a\tau} c_{i\sigma}^{b\tau'}. 
\end{eqnarray}
Technically, the new fields are embedded  into the path integral for partition function by insertion of the functional $\delta$-functions. To this end, we introduce  the Nambu-basis $\Psi_i=(c_{i\uparrow}, \bar{c}_{i\downarrow})^T$, $\bar{\Psi}_i=(\bar{c}_{i\uparrow}, c_{i\downarrow})$, and the matrix Green's function 
\begin{equation}
\mathbf{G}_{\tau\tau'}=\left(\begin{array}{cc}
G_{\tau\tau'} & \bar{F}_{\tau\tau'} \\ 
F_{\tau\tau'} & -G_{\tau'\tau}
\end{array}\right).  
\end{equation}
Then the functional $\delta$-functions are enforced by the conjugated matrix field 
\begin{equation}
\mathbf{\Sigma}_{\tau\tau'}=\left(\begin{array}{cc}
\Sigma_{\tau\tau'} & \bar{\Xi}_{\tau\tau'} \\ 
\Xi_{\tau\tau'} & -\Sigma_{\tau'\tau}
\end{array}\right).  
\end{equation}
as follows 
\begin{equation}
{\mathbf 1}=\int [D\mathbf{\Sigma}, \mathbf{G}] \exp\left[\sum_{i=1}^N \bar{\Psi}_i \mathbf{\Sigma}\Psi_{i}
+N\mathrm{Tr}\left(\mathbf{\Sigma}\mathbf{G}\right)\right],
\end{equation}
The dual fields $\Xi$, $\bar{\Xi}$, and $\Sigma$ play the role of anomalous and normal self-energies respectively. 

Furthermore, we perform Hubbard-Stratonovich transformation to decouple the Hubbard term in the Cooper channel, introducing site-local complex fields $\Delta_i$,  
\begin{eqnarray}
\nonumber && 
\exp\left[U\sum_i\int d\tau \bar{c}_{i\uparrow}^{\tau} \bar{c}_{i\downarrow}^{\tau} c_{i\downarrow}^{\tau} c_{i\uparrow}^{\tau}\right]
\\ 
\nonumber &&
=\int[D\bar{\Delta}_i^{\tau},\Delta_i^{\tau}]\exp\left[-\frac{1}{U}\sum_i\int d\tau\left(\bar{\Delta}^{\tau}_i\Delta^{\tau}_i\right)  \right.  \\ 
&& \left. 
+\sum_i\int d\tau \left(\bar{\Delta}^{\tau}_i c_{i\downarrow}^{\tau} c_{i\uparrow}^{\tau} + \Delta^{\tau}_i  \bar{c}_{i\uparrow}^{\tau} \bar{c}_{i\downarrow}^{\tau}  \right)
\right]. 
\label{decoupleHubbard}
\end{eqnarray}
After the decoupling procedure, the action reads: 
\begin{eqnarray}
\nonumber && 
S=\sum_{i=1}^N\int_0^{\beta} d\tau d\tau'\left\{\sum_{\sigma=\uparrow,\downarrow} \bar{c}_{i\sigma}^{\tau'}\left(\delta_{\tau'\tau}\partial_{\tau}+\Sigma_{\tau'\tau}\right) c_{i\sigma} 
 \right. \\ 
\nonumber && 
+\left. 
\Xi_{\tau'\tau}\bar{c}_{i\uparrow}^{\tau'} \bar{c}^{\tau}_{i\downarrow}  +  \bar{\Xi}_{\tau'\tau} c_{i\downarrow}^{\tau'}  c^{\tau}_{i\uparrow}+ (\bar{\Delta}_i^{\tau}c_{i\downarrow}^{\tau}  c_{i\uparrow}^{\tau} + \Delta_i^{\tau}\bar{c}_{i\uparrow}^{\tau}\bar{c}_{i\downarrow}^{\tau})\delta_{\tau\tau'}\right\}\\ 
\nonumber &&    
-N \int_0^{\beta} d\tau d\tau' \left\{2\Sigma_{\tau'\tau}G_{\tau\tau'}+\Xi_{\tau'\tau}\bar{F}_{\tau\tau'} +  \bar{\Xi}_{\tau'\tau} F_{\tau\tau'} \right. \\
\nonumber && 
\left. 
+\frac{J^2}{64} \left[ \bar{F}_{\tau\tau'}^2F_{\tau\tau'}^2  +
G_{\tau\tau'}^4\right] \right\}+
\frac{1}{U}\sum_{i=1}^N\int_0^{\beta}d\tau \bar{\Delta}_i^{\tau}\Delta_i^{\tau}.  \\ 
\label{S_general}
\end{eqnarray}

\subsection{Saddle point ansatz} 
We assume the fields $\Delta_i$ to be time- and site-independent at the saddle point. 
Then we integrate out fermion fields, and obtain the action in the form 
\begin{eqnarray}
\nonumber &&
\frac{S}{N}=-\sum_{\omega_n}\ln\left[(-i\omega_n+\Sigma(\omega_n))^2\right. \\ 
\nonumber && 
-\left. 
(\bar{\Xi}(\omega_n)+\bar{\Delta})(\Xi(\omega_n)+\Delta) \right]+
\frac{\beta}{U}\bar{\Delta}\Delta   
\\ 
\nonumber  && 
-\int d\tau d\tau'\left( \Xi_{\tau'\tau}\bar{F}_{\tau\tau'}+
\bar{\Xi}_{\tau'\tau}F_{\tau\tau'}+
2\Sigma_{\tau'\tau}G_{\tau\tau'}\right) \\ 
&& 
-\frac{J^2}{64}\int d\tau d\tau'\left[G_{\tau\tau'}^4+
\bar{F}_{\tau\tau'}^2  F_{\tau\tau'}^2 \right].
\label{ActionSP}
\end{eqnarray}
Variation of the action Eq. (\ref{ActionSP}) results in the following set of saddle point equations 
\begin{eqnarray}
\nonumber && 
\hskip -.4cm \frac{\Delta}{U}\!=\!T \sum_{\omega_n} \frac{\Delta+\Xi(\omega_n)}{(\omega_n+i\Sigma(\omega_n))^2+(\bar{\Xi}(\omega_n)+\bar{\Delta})(\Xi(\omega_n)+\Delta)}, \\ 
&& \label{DeltaSP}\\
\nonumber && 
\hskip -.4cm F(\omega_n)\!=\!\frac{-(\Delta+\Xi(\omega_n))}{(\omega_n+i\Sigma(\omega_n))^2+(\bar{\Xi}(\omega_n)+\bar{\Delta})(\Xi(\omega_n)+\Delta)}, \\ 
&& \label{F_SP}\\
 &&   \label{eq:A-Xi}
\hskip -.4cm \Xi_{\tau\tau'}=-\frac{J^2}{32}\bar{F}_{\tau\tau'}F_{\tau\tau'}^2,  \label{Xi_SP} \\
\nonumber && 
\hskip -.3cm G(\omega_n)\!=\!\frac{-i\omega_n+\Sigma(\omega_n)}{(\omega_n+i\Sigma(\omega_n))^2+(\bar{\Xi}(\omega_n)+\bar{\Delta})(\Xi(\omega_n)+\Delta)},  \\
&& \label{G_SP} \\
&& 
\hskip -.4cm \Sigma_{\tau\tau'}=\frac{J^2}{32} G_{\tau\tau'}^3. \label{Sigma_SP}
\end{eqnarray}
Note the relation 
\begin{equation}
\frac{\Delta}{U}=- T \sum_{\omega_n} F(\omega_n)=- F_{\tau\tau}. 
\label{DeltatoF}
\end{equation}
Hereafter we restrict ourselves to the case of the half-filling, where, due to the particle-hole symmetry, the normal components are odd, while anomalous are even functions of time, eg., $\Xi_{\tau'\tau}=\Xi_{\tau\tau'}$,  $\Sigma_{\tau'\tau}=-\Sigma_{\tau\tau'}$.

\subsection{Approximate solution of the mean-field equations}

The anomalous fields $\Xi$ and $F$, entering the saddle point equations, admit non-zero solutions only in the presence of $\Delta$. Similarly to the 
BCS case, we'll find that $F\propto \Delta$. According to Eq.~(\ref{eq:A-Xi}), $\Xi\propto F^3\propto \Delta^3$. Therefore in the limit of (exponentially) small 
$\Delta$ one may consider dropping $\Xi$ from the set of the mean-field equations and restricting them down to:  
\begin{eqnarray}
&&   
 G(\omega_n)=\frac{-i\omega_n+\Sigma(\omega_n)}{(\omega_n+i\Sigma(\omega_n))^2+\Delta^2}, \label{SYK_G}\\ 
&&
\Sigma_{\tau\tau'}=\frac{J^2}{32} G_{\tau\tau'}^3, \label{SYK_Sigma}\\ 
&& 
F(\omega_n)= -\frac{\Delta}{(\omega_n+i\Sigma(\omega_n))^2 +\Delta^2}, 
\label{F_approx}
\end{eqnarray} 
where we fixed the phase of $\Delta$ to make the latter real. We'll see below that neglecting $\Xi$ is not, strictly speaking, justified, even for the small  $\Delta$.
Nevertheless Eqs.~(\ref{SYK_G})--(\ref{F_approx}) will be shown to be a qualitatively (if not quantitatively) accurate representation of the full set. 
Eqs.~(\ref{SYK_G}), (\ref{SYK_Sigma}) are  the known saddle point equations of the SYK model, modified by the presence of a finite $\Delta$. 
In the normal phase ($\Delta=0$) Eqs. (\ref{SYK_G}), (\ref{SYK_Sigma}) exhibit an approximate conformal invariance at long times. 
Their solutions  behave like $G(\tau)\sim \mathrm{sign}(\tau)/\sqrt{J|\tau|}$ and $\Sigma(\tau)\sim \mathrm{sign}(\tau) \sqrt{J}/|\tau|^{3/2}$. 
 Assuming for a moment that 
$\Sigma\gg \Delta, \omega_n$, one finds $F(\omega_n)\propto \Delta/(J|\omega_n|)$. In the time representation this amounts to $F(\tau)\propto 
(\Delta/J)\ln(\tau_\Delta/\tau) $, where $\tau_\Delta$ is a long time cutoff to be discussed momentarily.   

A finite $\Delta$ creates a gap in the many-body spectrum, forcing the exponential decay of the correlation functions at a long imaginary time. We denote the corresponding time scale, given by the inverse of the energy gap, as $\tau_\Delta$. Following Ref.~\cite{esterlis2019cooper,hauck2019eliashberg}, based on these considerations we adopt the following variational ansatz for the normal and anomalous Green functions: 
\begin{eqnarray}      \label{eq:A-G}
G(\tau)&=&
-\frac{e^{-|\tau|/\tau_{\Delta}  }}{\sqrt{2\pi\tilde{J}|\tau|}}\, \mathrm{sgn}(\tau); \\
F(\tau)&=& -\frac{\Delta}{\pi \tilde J}\,  e^{-|\tau|/\tau_{\Delta}  } \ln\left(1+c\,\frac{\tau_\Delta}{|\tau|}\right),
                              \label{eq:A-F}
\end{eqnarray}
where $\tilde{J}=J/(4\sqrt{2\pi})$ and parameters  $\tau_\Delta$ and  $c$ are to be determined to satisfy Eqs.~~(\ref{SYK_G}), (\ref{F_approx}) in the limit of  small frequencies.   

To execute this program we first perform the Fourier transforms of  $G(\tau)$ and $\Sigma(\tau)=J^2G^3(\tau)/32$, finding: 
\begin{equation}  \label{G_largeOmega}
G(\omega_n)= \frac{  \mathrm{sgn}(\omega_n)}{i \sqrt{ \tilde{J} |\omega_n|} }; \quad \quad
\Sigma(\omega_n)=-i \sqrt{\tilde{J} |\omega_n| }\,  \mathrm{sgn}(\omega_n), 
\end{equation}
for $\omega\tau_\Delta\gg 1$ and 
\begin{equation}
G(\omega_n)= \frac{  \tau_{\Delta}^{3/2}\, \omega_n}{i \sqrt{2\tilde{J}} };\quad\quad  
                                                                        \label{G_smallOmega} 
\Sigma(\omega_n)=-i \frac{\sqrt{\tilde{J}\tau_{\Delta}} \, \omega_n }{\sqrt{6}} , 
\end{equation}
for $\omega\tau_\Delta\ll 1$. One notices that in both limits $\Sigma(\omega_n)\gg \omega_n$ and therefore the latter may be neglected in Eqs.~~(\ref{SYK_G}), (\ref{F_approx}). In the limit $\omega\tau_\Delta\gg 1$ one also notices that $\Sigma(\omega_n)\gg \Delta$ and thus 
$G(\omega_n)= -1/\Sigma(\omega_n)$, which is consistent with Eq.~(\ref{G_largeOmega}). This consistency is a consequence of our choice of 
$\tilde J$.  In the opposite limit, $\omega\tau_\Delta\ll 1$,   $\Sigma(\omega_n)\ll \Delta$ and thus 
$G(\omega_n)= \Sigma(\omega_n)/\Delta^2$. Combining this with Eq.~(\ref{G_smallOmega}), one finds for the inverse energy gap
\begin{equation}
\tau_{\Delta}=\frac{\tilde{J}}{\sqrt{3}\, \Delta^2}. 
\label{tau_Delta}
\end{equation}
Notice that the gap scales as $\Delta^2/J\ll \Delta$. This is a consequence of the superconductivity being formed from the non Fermi liquid normal state.

We turn now to the anomalous function. According to Eqs.~(\ref{F_approx}) and (\ref{G_largeOmega}) its high energy limit is given by:
\begin{equation}
F(\omega_n)=\frac{\Delta}{\Sigma^2(\omega_n)}=  -\frac{\Delta}{\tilde{J} |\omega_n| }.  
\label{F_LargeOmega}
\end{equation}
It's Fourier transform is $F(\tau)=-(\Delta/\pi \tilde J)\ln(\tau_\Delta/\tau)$, where $\tau_\Delta$ is adopted as a long time cutoff. This is exactly the variational 
form, Eq.~(\ref{eq:A-F}) at  $\tau\ll\tau_\Delta$. Finally to fix the constant $c$ in Eq.~(\ref{eq:A-F}) we demand the correct asymptotic at $\omega_n\to 0$, which is, according to Eqs.~(\ref{F_approx}) and (\ref{G_smallOmega}), $F(\omega_n=0)= \int\! d\tau F(\tau)=-1/\Delta$. Integrating Eq.~(\ref{eq:A-F}) with
$\tau_\Delta$ given by Eq.~(\ref{tau_Delta}),  one finds $c=7.58$. 

Finally, we can self-consistently  determine $\Delta$ using Eq.~(\ref{DeltatoF}). To this end one needs the anomalous function at the coinciding times: $F_{\tau\tau}=F(\tau=0)$. Putting UV cutoff instead, $\tau\sim 1/\tilde J$, one finds  
 \begin{equation}
\frac{\Delta}{U} =  \frac{\Delta}{\pi \tilde{J}}\, \ln\!\left(\tilde{J}\tau_\Delta \right) =  \frac{2\Delta}{\pi \tilde{J}}\, \ln\!\left(\frac{\tilde{J}}{\Delta} \right), 
\end{equation}
where the coefficient inside the logarithm is somewhat arbitrary. As a result, one finds 
\begin{equation}
\Delta\sim\tilde{J}\,  e^{-\frac{\pi \tilde{J}}{2 U}}.
\label{DeltasmallU}
\end{equation}
We conclude that, within  the mean-field treatment, the superconducting order parameter $\Delta$ is present at an infinitesimally small Hubbard attraction $U$. 

Let us now discuss the omission of the anomalous component of the self-energy, $\Xi(\omega_n)$, in Eqs.~(\ref{DeltaSP})--(\ref{Sigma_SP}). One expects that, since $\Xi\propto F^3$ and $F\propto \Delta\propto  e^{-\frac{\pi \tilde{J}}{2 U}}$, the anomalous self-energy is exponentially suppressed. In reality this is not  entirely the case. Indeed, let's evaluate $\Xi(\omega_n=0)\sim J^2 \int\! d\tau F^3(\tau)\sim J^2(\Delta/J)^3\tau_\Delta \sim\Delta$, where we have employed Eqs.~(\ref{eq:A-F}) and  (\ref{tau_Delta}). Therefore at small energies, $\omega_n\tau_\Delta\ll 1$,  $\Xi(\omega_n)\sim \Delta$, while for $\omega_n\tau_\Delta\gg 1$, $\Xi(\omega_n)\sim \Delta/(|\omega_n|\tau_\Delta)\propto \Delta^3$, as expected. Nevertheless, we observe that in the entire energy range 
$\Xi(\omega_n)\lesssim \Delta$ and therefore omitting $\Xi$ in Eqs.~(\ref{DeltaSP})--(\ref{Sigma_SP}) is not affecting the qualitative behavior of the Green functions, Eqs.~(\ref{eq:A-G}), (\ref{eq:A-F}), and the scaling of the inverse gap, Eq.~(\ref{tau_Delta}). It may affect, though, some of the numerical coefficients.   

In the opposite limit of large Hubbard coupling, $U>J$,  the spectral gap is of the order of $U$. Being the largest energy scale, the gap suppresses the SYK 
non Fermi liquid regime. This leads to $|\omega_n| \gg \Sigma(\omega_n), \Xi(\omega_n)$ and thus   Eq. (\ref{DeltatoF}) yields: 
\begin{equation}
\Delta=U\int \frac{d\omega}{2\pi}\frac{\Delta}{\omega^2+\Delta^2}=\frac{U}{2}. 
\label{Delta_LargeU}
\end{equation}

\section{Interaction constant in the quantum Kuramoto action}
\label{app:B}

\begin{figure}[ht]
	{\includegraphics[width=.5\textwidth]{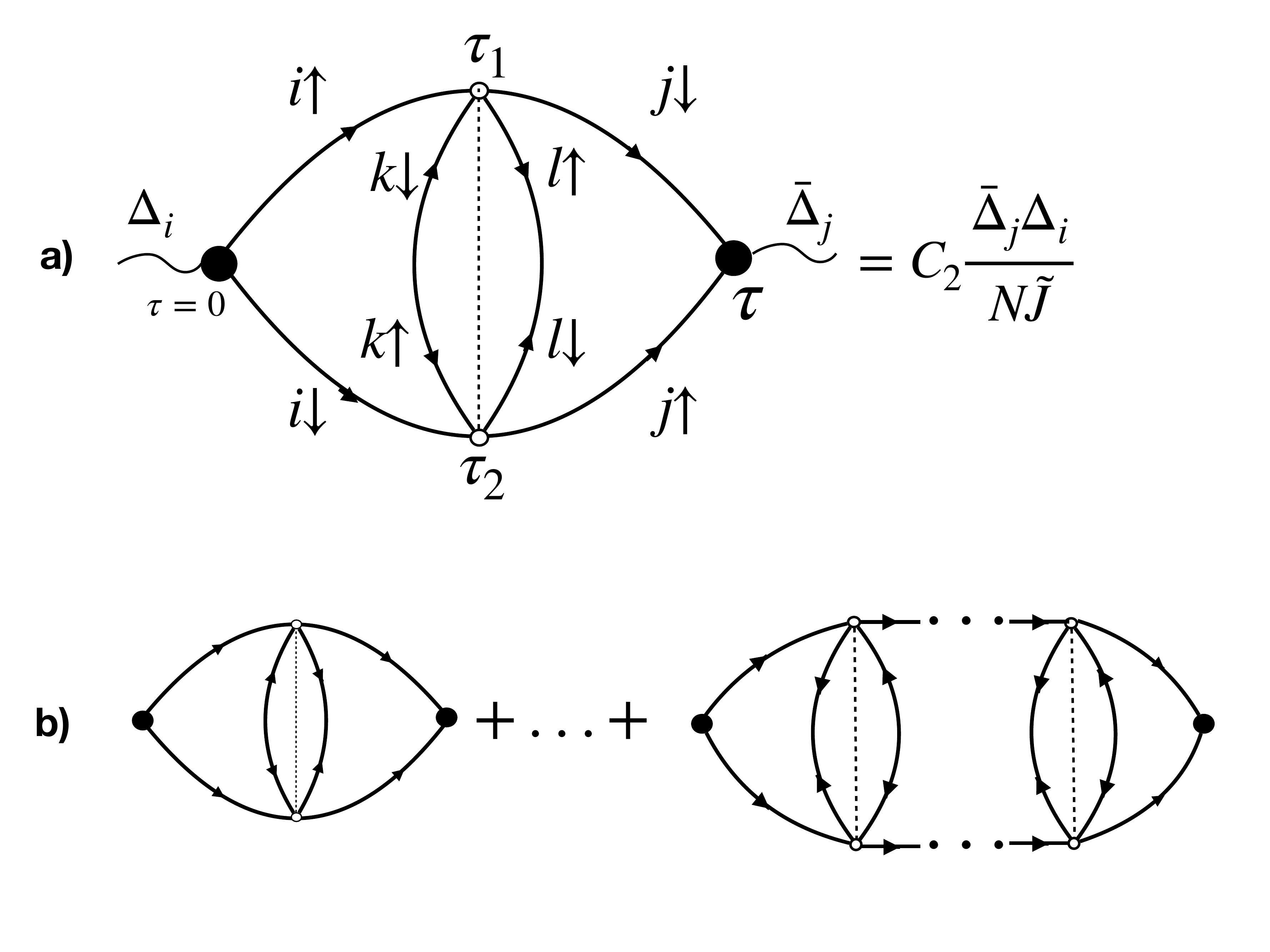}}
	\caption{ Diagrammatic representation of the off-diagonal Cooper susceptibility:  a) the lowest order diagram;  
		b) the ladder series. 
	}
	\label{FigDeltaLadder}
\end{figure}

Here we  derive  the interaction term for the phase fluctuations of the local order parameters on different sites, Eq. (\ref{eq:Kuramoto}). 
As explained below Eq. (\ref{eq:Kuramoto}), the corresponding coupling constant is proportional to the off-diagonal susceptibility to variations of the local order 
parameter, $\kappa_{ij} =\partial^2 E_\mathrm{GS}/\partial \Delta_i\partial \bar\Delta_j$. We thus consider the order parameters, $\Delta_i$, to be externally 
applied (proximitized) through an extra term in the Hamiltonian, $\sum_i \bar\Delta_i c_{i\downarrow} c_{i\uparrow}+h.c.$, and evaluate an induced 
energy change. Diagrammatically the latter is given by the order $1/N$ diagrams, Fig. \ref{FigDeltaLadder},  which involve normal and anomalous Green's functions, as well as the paired interaction vertices $\langle J_{ik;lj}^2\rangle =J^2/(4N)^3$. 

Since all propagators  are site-diagonal, correlations between distinct sites appear in the order  $1/N$ in expansion of the action. The physical mechanism of correlations between the superconducting fluctuations at different sites consists of correlated hopping of Cooper pairs facilitated by SYK-interactions $J_{ik;lj} c^{\dagger}_{i\sigma}c^{\dagger}_{k\sigma'}c_{l,\sigma'}c_{j,\sigma}$. Because all four sites here are distinct, no direct hopping of a Cooper pair is possible. Rather, a transfer of a single Cooper pair from a site $i$ to another site $j$ involves at least two correlated acts of interaction that cause transfer of two Cooper pairs from the sites $i, k$ to the sites $\ell, j$. Thereby the second Cooper pair plays the role of an assisting agent for the hopping of the first one.  The amplitude of an elementary jump of a Cooper pair from a site $i$ to a different site $j$,  assisted by a hopping of another Cooper pair from a site $k$ to a site $\ell$  is represented  by the diagram in Fig. \ref{FigDeltaLadder}a). The hopping of the assisting Cooper pair is depicted by  the insertion of an anomalous loop between the normal Green's functions. Thus, insertion of anomalous loops is a necessary ingredient of diagrammatic representation of interaction between superconducting fluctuations at different sites. 

Taking into account summation over the spin indexes and over the intermediate sites $k, \ell$, one obtains contribution  to the average susceptibility, $\kappa_{ij}^{(1)}$, from the lowest order diagram   in Fig. \ref{FigDeltaLadder}a):    
\begin{eqnarray}
\nonumber && 
\hskip -.3cm \kappa_{ij}^{(1)}=\frac{J^2}{16N}\int\! d\tau d\tau_1 d\tau_2\,  G(\tau_1)G(\tau_2)\\ 
&& 
\hskip -.3cm \times \bar{F}(\tau_1-\tau_2)F(\tau_1-\tau_2)    G(\tau- \tau_1)G(\tau-\tau_2) . 
\label{g1_general}
\end{eqnarray}
Substituting the variational solutions, Eqs. (\ref{eq:A-G}), (\ref{eq:A-F}), for normal and anomalous propagators and introducing dimensionless time-variables $t=\tau/\tau_{\Delta}$, $t_{1,2}=\tau_{1,2}/\tau_{\Delta}$, one finds 
\begin{eqnarray}
				\label{eq:A-kappa}
\kappa_{ij}^{(1)}= \frac{C_2^{(1)}}{N\tilde{J}} ,  
\end{eqnarray}
where  $C_2^{(1)}$ is given by a convergent integral, which does not depend on any parameters,   
\begin{eqnarray}
\nonumber && 
\hskip -.3cm C_2^{(1)}=\frac{1}{2\sqrt{3}\pi^3}\int\! dt dt_1 dt_2\,  \mathrm{sign}(t_1)\mathrm{sign}(t_2)\\ 
\nonumber && 
\hskip -.3cm \mathrm{sign}(t-t_1)\mathrm{sign}(t-t_2)
 \ln^2\left(1+\frac{7.58}{|t_1-t_2|}\right)\times  \\ 
\nonumber && 
\hskip -.3cm \frac{\exp\left[-|t_1|-|t_2|-|t-t_1|-|t-t_2|-2|t_1-t_2|\right]}{\sqrt{|t_1| |t_2| |t-t_1| |t-t_2|}}=1.58.   
\end{eqnarray}
This numerical constant should not be taken too seriously. Indeed, our variational ansatz for the propagators, Eqs. (\ref{eq:A-G}), (\ref{eq:A-F}), is not exact but only 
interpolates between correct short and long time asymptotics. The reason of presenting this calculation is to point out the absence of logarithmic factors. The latter  may be naively expected, due to the presence of two runs of the Cooper ladder in the diagram of Fig. \ref{FigDeltaLadder}a). If the anomalous loop in the middle would be confined in time to some scale $\tau_0\ll \tau_\Delta$, the diagram would be $\propto \ln^2(\tau_\Delta/\tau_0)\gg 1$. This is because the two integrals over $\tau$ and $\tau_1\approx\tau_2$ would be logarithmic. In this case summation of the entire Cooper ladder, Fig. \ref{FigDeltaLadder}b), 
would be of a crucial importance. However, our case happens to be different. The reason is that the anomalous loop has the same characteristic time scale, $\tau_\Delta$, as the normal Green functions, which form runs of the Cooper ladder. As a result, logarithms are not present and all the terms of the ladder have the same order 
of magnitude as the first diagram, Fig. \ref{FigDeltaLadder}a). Therefore the ladder summation  only changes the numerical coefficient, $C_2$, rather than 
the large logarithmic factor. Let us note in passing that Hubbard $U$, being time-local, induces the conventional logarithmic Cooper ladder and thus Eq.~(\ref{DeltasmallU}). This ladder, however, is strictly diagonal in the site index (and is already incorporated in diagonal $G$ and $F$, Eqs. (\ref{eq:A-G}), (\ref{eq:A-F})). The off-diagonal ladder and thus the off-diagonal susceptibility, $\kappa_{ij}$, needed  in the quantum Kuramoto action, requires long-range anomalous loops inserted in each run of the ladder as in Fig. \ref{FigDeltaLadder}b).

Another consequence of the long-range nature of the anomalous loop is that the susceptibility, $\kappa_{ij}$, Eq.~(\ref{eq:A-kappa}), is not proportional to $\Delta$, despite each of the two anomalous propagators, $F$, being proportional to $\Delta$, Eq.~(\ref{eq:A-F}). The reason is that the integrations range, given by $\tau_\Delta$, is inversely proportional to $\Delta^2$, Eq.~(\ref{tau_Delta}). In the absence of other long time cutoffs, e.g. a finite size gap, this leads to $\Delta$-independent susceptibility, Eq.~(\ref{eq:A-kappa}).

Finally, the interaction term in the quantum Kuramoto action, Eq.~(\ref{eq:Kuramoto}), is given by $\kappa_{ij}\Delta_i\bar\Delta_j+h.c.\sim
\kappa_{ij}|\Delta|^2 \cos(\phi_i-\phi_j)$. As a result, the interaction constant, $g$,  in the quantum Kuramoto action, Eq. (\ref{eq:Kuramoto}), is given by  
\begin{eqnarray}
g=C_2\, \frac{\Delta^2 }{\tilde{J}} \propto \tilde{J} e^{-\frac{\pi \tilde{J}}{U}}, 
\label{g_Kuramoto}
\end{eqnarray}
where constant $C_2\sim O(1)$ remains undetermined by these considerations. The fact that $g$ is linearly proportional to the {\em energy gap}, Eq.~(\ref{tau_Delta}),  (both being $\sim \Delta^2$) is analogous to the conventional $T=0$ Josephson energy.

\section{Richardson model and its generalizations}
\label{app:Richardson} 
In this Appendix we review some general aspects of the Richardson model and
its generalizations for  completeness.

\subsection{Richardson model}

The truncated BCS-like Richardson model of superconductivity \cite{Richardson1963}
involves some number
of doubly degenerated fermionic levels with the set of energies $\epsilon_{j}/2$, where $j=1,\dots, N$.
It describes the system
with  a fixed number, $M\leq N$, of the Cooper pairs.
It is assumed that  several energy levels are populated by  Cooper pairs while
levels with the single fermions are blocked.
The Hamiltonian reads
as
\begin{equation}
H_{R}= \frac{1}{2}\sum_{j,\sigma= \uparrow\downarrow}^N \epsilon_{j} c^{\dagger}_{j\sigma}c_{j\sigma} -
G\sum_{jk}c^{\dagger}_{j\uparrow}c^{\dagger}_{j \downarrow}c_{k \downarrow}c_{k\uparrow}, 
\end{equation}
where $c^{\dagger}_{j\sigma}$ are the fermion operators and $G$ is a
coupling constant providing the attraction between fermions. 
In terms of the hard-core boson operators
it reads as
\begin{equation}
H_{R}=\sum_j\epsilon_j b^{\dagger}_jb_j - G \sum_{jk}b^{\dagger}_jb_k,
\end{equation}
where
\begin{equation}
[b^{\dagger}_j,b_k]= \delta_{jk}(2N_j-1), \qquad b_j=c_{j \downarrow}c_{j \uparrow},\qquad N_j= b^{\dagger}_jb_j.
\end{equation}

The eigenfunctions of the Hamiltonian can be written as
\begin{equation}
|M\rangle=\prod_i^M B_i^\dagger  |\mathrm{vac}\rangle, \qquad B_i^\dagger \equiv \sum _j^N \frac{b^{\dagger}_j}{\epsilon_j- E_i}, 
\end{equation}
provided the set of energies $E_i$, where $i=1,\ldots, M$ satisfies the Bethe Anzatz (BA) equations
\begin{equation}
\label{BA}
G^{-1}= - \sum _j^N \frac{2}{\epsilon_j- E_i} + \sum _j^M \frac{1}{E_j- E_i}. 
\end{equation}
The many-body energies of the corresponding states read as
\begin{equation}
E(M)= \sum_i^M E_i. 
\end{equation}
For  nontrivial degeneracies of the energy levels, $d_j$, 
the BA equations read as
\begin{equation}
G^{-1}= - \sum _j^N \frac{d_j}{\epsilon_j- E_i} + \sum _j^M \frac{1}{E_j- E_i}. 
\end{equation}

It is convenient
to introduce the  pseudospin $Sl(2,R)$ algebra in   
terms of the creation-annihilation
operators for the Cooper pairs
\begin{equation}
t^{-}_j=b_j \qquad t^{+}_j=b^{\dagger}_j \qquad t^{0}_j=N_j-1/2.
\end{equation}
The Richardson Hamiltonian commutes with
the set of operators $R_j$ \cite{rivas}
\begin{equation}
R_i= -t^0_i -2G\sum^N_{i\neq j}\frac{\sum_{a=\pm,0}t_i^a t_j^a}{\epsilon_i -\epsilon_j},
\end{equation}
which are identified as the Gaudin Hamiltonians 
\begin{equation}
[H_{R},R_j]=[R_i,R_j]=0.
\end{equation}
Moreover the Richardson Hamiltonian itself 
can be expressed in terms of the operators $R_i$ as
\begin{equation}
H_{R}= \sum_j \epsilon_i R_i + G \left(\sum_i R_i\right)^2 + \mathrm{const}.
\end{equation}

The number of  orbitals, $N$, coincides
with a number
of sites in the Gaudin model and a coupling constant in the
Richardson Hamiltonian corresponds to the "boundary twist" 
in the Gaudin model. 
The commuting operators, $R_i$, are the residues of
the transfer matrix of the inhomogeneous twisted XXX spin 
chain in the semi-classical limit taken at inhomogeneities, $\epsilon_i$.
The BA equations for the Richardson model, Eq.~ (\ref{BA}) and for the Gaudin model exactly coincide.
The Richardson model can be described in terms
of  the conformal field theory, where the Cooper pairs correspond
to  screening operators \cite{sierraconf}.

\subsection{Russian Doll (RD) model  and  twisted 
inhomogeneous XXX spin chains}

A generalization of the Richardson model -
the so-called RD model \cite{lrs}, involves TRI breaking parameter,
$\alpha$. It's Hamiltonian is given by 
\begin{equation}
H_{\mathrm{RD}}= 2\sum_i^N (\epsilon_i- G)N_i -\bar{G}\sum_{j<k}
(e^{i\alpha} b^{\dagger}_k b_j +e^{-i\alpha} b^{\dagger}_jb_k).
\end{equation}
The  two   parameters $G, \bar G$ can be related to $\alpha$ as 
\begin{equation}
\alpha= \mathrm{arctanh}\left(\frac{\eta}{G}\right),
\end{equation}
where $\eta =\sqrt{\bar G^2 -G^2}$. 
It is  also useful to consider dimensionless parameters $g,\theta$
defined as $G=g d$ and $\eta =\theta d$, where $d$  is a mean value of  $(\epsilon_{i+1}-\epsilon_i)$ sequence.
The RD model reduces to the Richardson model in the limit $\eta\rightarrow 0$.

The RD model turns out to be integrable as well. Now instead
of the Gaudin model a proper spin chain counterpart
is  the generic quantum twisted inhomogeneous XXX spin chain \cite{rdsolution}. 
The equation defining a spectrum of the RD model reads as
\begin{equation}
\exp(-2i\alpha)\prod^N \frac{E_j-\epsilon_k -i\eta/2}
 {E_j-\epsilon_k + i\eta/2}=
 \prod^M \frac{\epsilon_j-\epsilon_k -i\eta}
 {\epsilon_j-\epsilon_k + i\eta}
\end{equation}
and coincides with the BA equations for the spin chain.
It reduces to the BA equation of the Richardson model
(\ref{BA})  in the limit
 $\eta \rightarrow 0$.

The  RD model enjoys  the gap equation, which reads as follows: 
\begin{equation}
\tilde{\Delta}_j= \sum_{i\neq j} V_{ij} \frac{\tilde{\Delta}_i}{\sqrt{(\epsilon_i - V_{ii})^2 + |\Delta_i|^2}},
\qquad \tilde{\Delta}_j =\Delta_j e^{i\phi_i}, 
\end{equation}
where $V_{ij}$ is a scattering potential, which depend on the parameters $G,\alpha$. In the thermodynamical 
limit it becomes an integral equation with  multiple solutions for the gaps.
Different solutions to the gap equation  yield different  superconducting states.

Solutions of the gap equation in the large $N$ limit are parametrized as follows:
\begin{equation}
 \Delta_n= \frac{\omega}{\sinh t_n}, \quad t_n= t_0+ \frac{\pi n }{\theta} \quad n=0,1 \dots, 
\end{equation}
 where $t_0$ is a solution to the following equation:
\begin{equation}
  \tan(\theta t_0)= \frac{\theta}{g}\qquad 0<t_0< \frac{\pi}{\theta}
\end{equation}
and $\omega =dN$ for  equal  spacing $(\epsilon_{i+1}-\epsilon_i) = \mathrm{const}$. This
behavior can be derived in the mean-field approximation \cite{lrs}.
In the limit $\theta \rightarrow 0$
the gaps $\Delta_{n>0} \rightarrow 0$ and
\begin{equation}
t_0=\frac{1}{g},\qquad \Delta_0= 2\omega e^{-\frac{1}{g}}.
\end{equation}
This way the standard BCS expression for the gap is recovered. At  a weak coupling
the gaps behave as
\begin{equation}
\Delta_n \propto \Delta_0 e^{-\frac{n\pi}{\theta}}. 
\end{equation}

For Cooper pair degeneracies on  orbitals,  $d_i$,  the
RD model is modified a bit and is related to the
higher spin XXX spin chain. The local spins $s_i$ are determined
by the corresponding pair degeneracy, $d_i$, of the i-th orbital
\begin{equation}
s_i=d_i/2
\end{equation}
and the corresponding BA equations read as 
\begin{equation}
\exp(-2i\alpha)\prod^N \frac{E_j-\epsilon_k -i\eta/2+ i\eta s_i}
 {E_j-\epsilon_k + i\eta/2 - i\eta s_i}=
 \prod^M \frac{\epsilon_j-\epsilon_k -i\eta}
 {\epsilon_j-\epsilon_k + i\eta}.
\end{equation}

The RD model involves an interesting RG behavior 
of couplings with respect to RG time $s=\log N$,  \cite{lrs}. The 
coupling constant exhibit the cyclic RG flow (a recent review on the cyclic RG can be found in \cite{gorsky}), 
while the TRS parameter does not renormalize 
\begin{equation}
 g_{N-1}=g_N + \frac{1}{N}(g_N^2 + \theta^2),\qquad \theta_{N-1}=\theta_N
\end{equation}
\begin{equation}
 g(s+ \lambda) =g(s),\qquad g(e^{-\lambda}N)=g(N).
 \end{equation}
 The RG period reads as
\begin{equation}
 \lambda=\frac{\pi}{\theta}
 \end{equation}
and the total number of the independent gaps in the model is
\begin{equation}
N_{\mathrm{gaps}}\propto \frac{\theta}{\pi} \log N.
\end{equation}
The cyclic RG behavior reflects
the breaking of the scale invariance down to the discrete subgroup 
and the spectrum of gaps manifests in the Efimov scaling
\begin{equation}
\Delta_{n+1}=e^s\Delta_n
\end{equation}
The sizes of the Cooper
pairs in the n-th  condensates also have the
Efimov-like scaling.

\subsection{Possible generalizations}

Here we consider  generalizations of the Richardson model, involving four-boson
interactions. The Hamiltonian (\ref{eq:Richardson}), appropriate for large $U$, is 
\begin{equation}  \label{eq:C-Richardson}
H_{\mathrm{gR}}\propto -\sum_{ijkl}^N b_i^{\dagger}b_{j}^{\dagger}b_k b_l. 
\end{equation}
Hence one may question if Hamiltonians with  four-boson interactions 
can be derived from the commuting set, $R_i$. Such representation  would
prove the integrability of the model. It is known that the
Hamiltonians, $R_i$, obey a nontrivial algebraic relation \cite{dimo18}
\begin{equation}
R_{i}^2= G^2 + \sum_j \frac{R_j}{\epsilon_i -\epsilon_j} + \frac{3}{4} \sum_{i\neq j} \frac{1}{(\epsilon_i -\epsilon_j)^2} 
\end{equation}
which follows from the hidden algebraic structure of the Gaudin model.
Therefore $R_i^2$ yield  two-boson interaction term only. 

To obtain the four-boson interaction term we can consider the quadratic form 
\begin{equation}
\label{eq:C-H4}
H_{4}= \sum_{ij}A_{ij}(\epsilon_i) R_i R_j
\end{equation}
with arbitrary matrix, $A_{ij}$. The integrable Hamiltonians, $H_4$, involve the desired four-boson
interactions.  In general, if  $\epsilon_i\neq 0$, the resulting interaction
coupling constants are site- and $\epsilon_i$-dependent. In 
our case all $\epsilon_i=0$ and hence the Hamiltonian (\ref{eq:C-Richardson}) can be considered
as the peculiar limit of the generic quadratic form, Eq.~(\ref{eq:C-H4}). Moreover, 
all  Bethe states creation operators, $B_i$, at $\epsilon_i=0$ reduce to the single operator
$B_0= \sum_{i} b_{i}^{\dagger}$.

\nocite{*}

\bibliography{SYK_superconductor.bib}

\end{document}